\documentclass[useAMS]{mn2e}

\usepackage{amsmath}
\newcommand{\dfn}[3]{\dfrac{\mathrm{d}^{#3} #1}{\mathrm{d} #2^{#3}}}
\newcommand{\pdn}[3]{\dfrac{\partial^{#3} #1}{\partial #2^{#3}}}
\newcommand{\rd}{\mathrm{d}}

\usepackage{subfigure}
\usepackage[dvips]{graphicx}

\def\pmb#1{\mbox{\boldmath$#1$}}
\def\pmbmt#1{\pmb{\sf #1}}

\def\be{\begin{eqnarray}}
\def\ee{\end{eqnarray}}

\def\rmi{\rm i}

\begin{document}
\title[non-axisymmetric magnetic modes]
  {Non-axisymmetric magnetic modes of neutron stars with purely poloidal magnetic fields}
\author[H. Asai, U. Lee, and S. Yoshida]
  {Hidetaka Asai, Umin Lee, and Shijun Yoshida \\
Astronomical Institute, Tohoku University, Sendai 980-8578, Japan}

\date{Received / Accepted}
\pubyear{0000} \volume{000} \pagerange{000-000} \onecolumn

\maketitle \label{firstpage}
 \nokeywords

\begin{abstract}
We calculate non-axisymmetric oscillations of neutron stars magnetized by
purely poloidal magnetic fields. We use polytropes of
index $n=1$ and 1.5 as a background model, where we ignore the equilibrium
deformation due to the magnetic field. 
Since separation of variables
is not possible for the oscillation of magnetized stars, we employ
finite series expansions for the perturbations using spherical
harmonic functions. Solving the oscillation equations as the boundary and
eigenvalue problem, we find two kinds of discrete magnetic modes, that is, stable (oscillatory) magnetic modes
and unstable (monotonically growing) magnetic modes.
For isentropic models,
the frequency or the growth rate of the magnetic modes is exactly
proportional to $B_{\rm S}$, the strength of the field at the surface.
The oscillation frequency and the growth rate are affected by the buoyant force in the interior,
and the stable stratification tends to stabilize the unstable magnetic modes.
\end{abstract}


\section{Introduction}
Neutron stars are believed to have strong magnetic fields. 
The field strength $B_S$ at the surface 
is estimated to be $\sim 10^{12}$ G for pulsars and $\sim 10^{15}$ G for magnetar
candidates, but we do not have any good knowledge of the configuration of magnetic fields
inside the star.
Recently, quasi-periodic oscillations (QPOs)
in the tail of the giant X/$\gamma$-ray flare were detected from the
soft $\gamma$ ray repeaters (SGRs) (e.g., Israel et al. 2005; Strohmayer \&
Watts 2005). SGRs belong to what we call
magnetars. Giant flares have so far been observed
only from three SGRs, that is, SGR 0526-66, 1900+14, and 1806-20, and
just once for each of the SGRs, indicating that giant flares in
magnetars are quite rare events. The QPOs are now regarded as a manifestation of global oscillations of
the underlying neutron stars, and it is expected that they can be used
for seismological studies of the magnetars. 
However, it is not an easy task to properly carry out modal analysis of magnetar
candidates because of their extremely strong magnetic fields, which can significantly
modify the modal property of the stars.
Because separation of variables between the radial and angular coordinates
is not possible for the perturbations in magnetized stars, we have to employ
series expansions to represent the perturbations in linear analysis, which possibly makes the analysis difficult,
particularly when singular points inherent to the governing equations appear in the interior of the star (see, e.g., Asai, Lee, \& Yoshida 2015).

Many authors have calculated axisymmetric ($m=0$) oscillation modes of magnetized stars having a purely
poloidal magnetic field, where $m$ is the azimuthal wave number of modes.
Using a toy model, Glampedakis, Samuelsson, and Andersson (2006) and Levin (2006,
2007) investigated global oscillation modes residing in the fluid core and in
the solid crust both threaded by a magnetic field, and they showed that frequency resonance
between modes in the core and in the crust could be important. 
Lee (2008) and Asai \& Lee (2014) carried out normal mode calculations of axisymmetric toroidal modes
and found discrete modes. 
van Hoven \& Levin (2011, 2012) suggested the existence of
discrete modes in the gaps between frequency continua, using
spectral method. 
Besides the normal mode analyses mentioned above, however,
most authors have used MHD
simulations to investigate the modal properties of magnetized stars (e.g., Cerd\'a-Dur\'an
et al. 2009; Colaiuda \& Kokkotas 2011; Gabler et al. 2011, 2012; Sotani et al. 2008). 
Sotani \& Kokkotas (2009) calculated axisymmetric polar-Alfv\'en oscillations
and found that continuous frequency spectra are not formed. 
Colaiuda \& Kokkotas (2012) calculated mixed polar and axial oscillation modes of
magnetized star having both poloidal and toroidal magnetic fields for
axisymmetric perturbations. 
They suggested that the oscillation spectra
can be significantly modified by the toroidal magnetic field
component. 
Gabler et al. (2013a) calculated magnetic oscillation modes
for various magnetic field configurations (e.g., purely poloidal, purely
toroidal, mixed poloidal and toroidal).

For non-axisymmtric ($m\not=0$) modes of magnetized stars, 
however, there are only a few numerical studies of global oscillations. 
For purely toroidal magnetic field configuration, Lander et al. (2010), Passamonti \& Lander (2013), and
Asai, Lee, and Yoshida (2015) calculated non-axisymmetric oscillation modes. 
Lander \& Jones (2011) calculated non-axisymmetric oscillation modes of magnetized rotating star with
purely poloidal magnetic field using MHD simulations, and they obtained polar-led
Alfv\'en modes, which reduce to inertial modes in the limit of ${\cal M}/{\cal T}\to 0$, where $\cal M$ and $\cal T$ are magnetic
and rotation energies of the star. 
It is important to note that Lander \& Jones (2011) suggested that the axial-led Alfv\'en modes
could be unstable.

In this paper, employing the normal mode analysis,
we calculate non-axisymmetric ($m\not=0$) oscillation modes of
magnetized stars having purely poloidal magnetic fields.
We assume that the gravitational energy dominates the magnetic energy so that
the stellar deformation due to the magnetic field can be safely neglected. 
No effect of rotation is considered. 
For normal mode analysis in this paper, we employ finite series expansions for perturbed quantities 
to derive a finite set of coupled linear ordinary differential equations, 
which is solved as the boundary and eigenvalue problem by imposing appropriate boundary conditions.
This paper is organized as follows. \S2 describes the method used to
construct a magnetized equilibrium stellar model, and perturbation
equations for non-axisymmetric oscillation modes in magnetized stars
are derived in \S3. Numerical results are summarized in \S4 and
discussions about magnetic modes are summarized in \S5. we
conclude in \S6. The details of the oscillation equations solved in this
paper and suitable boundary conditions imposed at the stellar center
and surface are given in Appendix.

\section{Equilibrium model}

Although we are concerned with magnetized neutron stars, in this study, we consider the problem within the frame work of Newtonian ideal 
magnetohydrodynamics for the sake of simplicity. 
For the modal analysis of magnetized stars composed of the infinitely conductive fluid, in this paper, magnetic fields in equilibrium 
are treated as perturbations about 
non-magnetized and non-rotating star. As mentioned before,  we ignore the deformation of the star due to the magnetic stress and use 
self-gravitating polytropic spheres for the matter distribution of the star. Then, the mass density and pressure are 
given by
 \begin{eqnarray}
 \rho = \rho_0 \Theta^n \,, \quad p = p_0 \Theta^{n+1} \,, 
 \end{eqnarray}
 where $\rho_0$ and $p_0$ are the density and pressure values at the center of the star, respectively, 
 and $n$ and $\Theta$ are the polytropic index and the Lane-Emden function, respectively.  
Imposed stationary axisymmetric magnetic fields are assumed to be the purely poloidal and dipole ones, given by
\begin{eqnarray}
B_r=2f(r)\cos\theta, \quad
B_\theta=-\left[r\dfn{f(r)}{r}{}+2f(r)\right]\sin\theta, \quad B_\phi=0,
\end{eqnarray}
in which $\nabla\cdot\pmb{B}=0$ is automatically satisfied. Here and henceforth, the spherical polar coordinate $(r,\theta,\phi)$ has been employed. 
The function $f(r)$ in Eq. (2) is
determined by the Ampere law, $\nabla\times\pmb{B}=4\pi j_\phi r\sin\theta\pmb{e}_\phi$, which leads to 
\begin{eqnarray}
\dfn{f}{r}{2}+\frac{4}{r}\dfn{f}{r}{}=-4\pi j_\phi,
\label{eq:grad}
\end{eqnarray}
where $j_\phi$ is related to the toroidal current and needs to satisfy integrability conditions for the ideal magnetohydrodynamic (MHD) equations. 
In this study, we assume that 
$j_\phi=c_0\rho$ where $c_0$ is a constant determined by the boundary condition at the surface of the star.  
Near the center of the star, the function $f$ behaves as 
\begin{eqnarray}
f=\alpha_0+{\cal{O}}(r^2),
\end{eqnarray}
where $\alpha_0$ is a constant determined by the boundary condition at the surface of the star.
We assume that $j_\phi=0$ outside the star. Thus, the exterior solution $f^{(\rm ex)}$ is given by $f^{(\rm  ex)}=\mu_b/r^3$, 
where $\mu_b$ is the magnetic dipole moment of the star. We determine the constants $\alpha_0$ and $c_0$ so
that the interior solutions $f$ and $\rd f/\rd r$ are continuously matched with the exterior solutions
$f^{(\rm ex)}$ and $\rd f^{(\rm ex)}/\rd r$ at the surface of the star.
\begin{figure}
\begin{center}
\resizebox{0.47\columnwidth}{!}{
\includegraphics{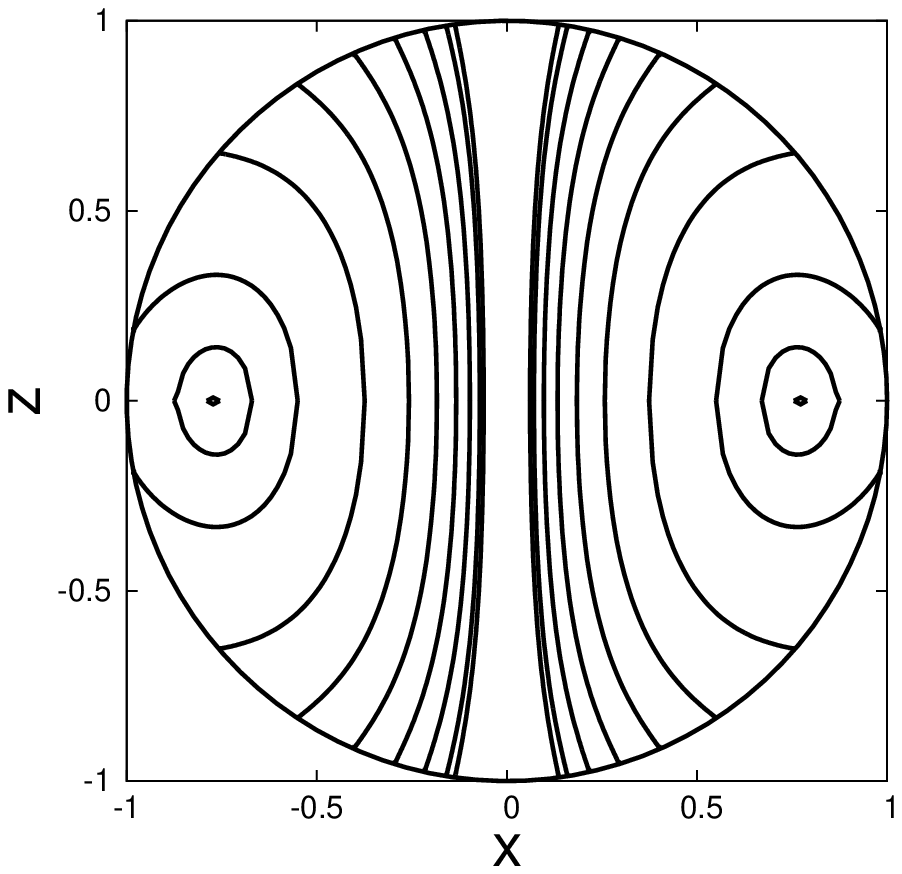}}
\hspace*{-1.2cm}
\resizebox{0.47\columnwidth}{!}{
\includegraphics{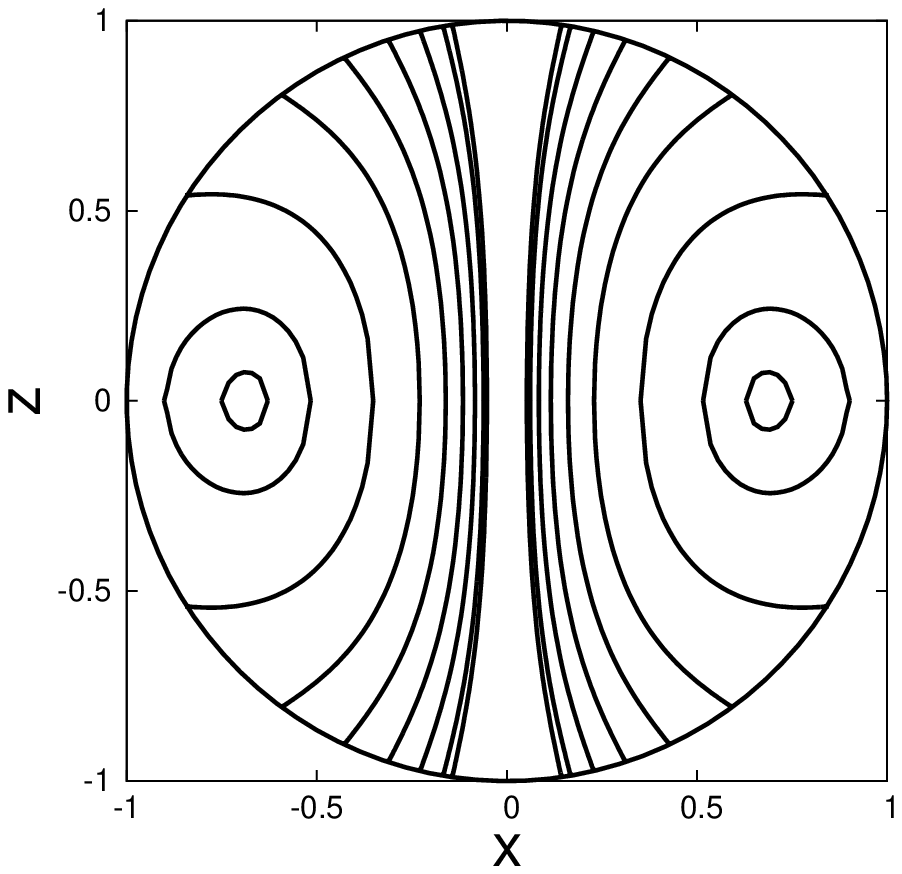}}
\end{center}
\caption{Magnetic field lines in the polytropic model of index $n=1$ (left panel)
and $n=1.5$ (right panel). }
\end{figure}
The magnetic field lines in the polytropes of index $n=1$ and 1.5 are shown in Figure 1.
We see that there exist field lines closed within the star.

\section{Perturbation equations}

The governing equations for non-radial oscillations of magnetized
stars are obtained by linearizing the ideal MHD equations. Since the
equilibrium state is stationary and axisymmetric, the time and azimuthal dependence of the
perturbed quantities is given by the factor $\exp[i(\sigma t+m\phi)]$,
where $m$ is the azimuthal wave number. 
The linearized basic
equations that govern the adiabatic, non-radial oscillations of 
magnetized stars are written as
\begin{eqnarray}
-\sigma^2\pmb{\xi}=-\nabla\Phi^\prime-\frac{1}{\rho}\nabla p^\prime+\frac{\rho^\prime}{\rho^2}\dfn{p}{r}{}\pmb{e}_r+\frac{1}{4\pi\rho}\left[\left(\nabla\times\pmb{B}^\prime\right)\times\pmb{B}+\left(\nabla\times\pmb{B}\right)\times\pmb{B}^\prime\right],
\label{eq:eqmot}
\end{eqnarray}
\begin{eqnarray}
\rho^\prime+\nabla\cdot\left(\rho\pmb{\xi}\right)=0,
\end{eqnarray}
\begin{eqnarray}
\frac{\rho^\prime}{\rho}=\frac{p^\prime}{\Gamma_1p}-\frac{\xi_r}{r}rA,
\label{eq:cont}
\end{eqnarray}
\begin{eqnarray}
\pmb{B}^\prime=\nabla\times\left(\pmb{\xi}\times\pmb{B}\right),
\end{eqnarray}
where $(^\prime)$
indicates the Eulerian perturbation, and $rA$ in equation (\ref{eq:cont}) denotes the Schwarzschild discriminant
defined as
\begin{eqnarray}
rA=\dfn{\ln\rho}{\ln r}{}-\frac{1}{\Gamma_1}\dfn{\ln p}{\ln r}{},
\end{eqnarray}
and $\Gamma_1=(\partial\ln p/\partial\ln\rho)_{\rm ad}$.
For polytropes of the index $n$, the adiabatic
exponent for the perturbations are assumed to be given by
\begin{eqnarray}
\frac{1}{\Gamma_1}=\frac{n}{n+1}+\gamma
\end{eqnarray}
with $\gamma$ being a constant, for which $rA=-\gamma(\rd\ln p/\rd\ln r)\equiv \gamma V$. 
The star may be called radiative for $\gamma<0$, isentropic for $\gamma=0$, and convective for $\gamma>0$.
For radiative stars, we have $g$-modes, whose oscillation frequency is proportional to $\sqrt{-\gamma}$.
For simplicity, we employ the Cowling approximation, neglecting $\Phi^\prime$.

Because of the Lorentz force term in equation (\ref{eq:eqmot}), separation of
variables for the perturbations is impossible between the radial
coordinate $r$ and the angular coordinate $\theta$. We
therefore expand the perturbations in terms of the spherical harmonic
functions $Y_l^m(\theta,\phi)$ with different $l$'s for a given
azimuthal index $m$. The displacement vector $\pmb{\xi}$ is given by
(see e.g., Lee 2005, 2007)
\begin{eqnarray}
\xi_r=\sum_{j=1}^{j_{\rm max}}rS_{l_j}(r)Y_{l_j}^m(\theta,\phi),
\end{eqnarray}
\begin{eqnarray}
\xi_\theta=\sum_{j=1}^{j_{\rm max}}\left[rH_{l_j}(r)\pdn{}{\theta}{}Y_{l_j}^m(\theta,\phi)-irT_{l_j^\prime}(r)\frac{1}{\sin\theta}\pdn{}{\phi}{}Y_{l_j^\prime}^m(\theta,\phi)\right],
\end{eqnarray}
\begin{eqnarray}
\xi_\phi=\sum_{j=1}^{j_{\rm max}}\left[rH_{l_j}(r)\frac{1}{\sin\theta}\pdn{}{\phi}{}Y_{l_j}^m(\theta,\phi)+irT_{l_j^\prime}(r)\pdn{}{\theta}{}Y_{l_j^\prime}^m(\theta,\phi)\right],
\end{eqnarray}
and the vector $\pmb{B}^\prime$ is given by
\begin{eqnarray}
B_r^\prime=\sum_{j=1}^{j_{\rm max}}rb^S_{l_j^\prime}(r)Y_{l_j^\prime}^m(\theta,\phi),
\end{eqnarray}
\begin{eqnarray}
B_\theta^\prime=\sum_{j=1}^{j_{\rm max}}\left[rb^H_{l_j^\prime}(r)\pdn{}{\theta}{}Y_{l_j^\prime}^m(\theta,\phi)-irb^T_{l_j}(r)\frac{1}{\sin\theta}\pdn{}{\phi}{}Y_{l_j}^m(\theta,\phi)\right],
\end{eqnarray}
\begin{eqnarray}
B_\phi^\prime=\sum_{j=1}^{j_{\rm max}}\left[rb^H_{l_j^\prime}(r)\frac{1}{\sin\theta}\pdn{}{\phi}{}Y_{l_j^\prime}^m(\theta,\phi)+irb^T_{l_j}(r)\pdn{}{\theta}{}Y_{l_j}^m(\theta,\phi)\right],
\end{eqnarray}
where $l_j=|m|+2(j-1)$ and $l_j^\prime=l_j+1$ for even modes, and
$l_j=|m|+2j-1$ and $l_j^\prime=l_j-1$ for odd modes, respectively, and
$j=1,2,3,...,j_{\rm max}$.
The Euler perturbations of the pressure and density are given by
\begin{eqnarray}
p^\prime=\sum_{j=1}^{j_{\rm max}}p_{l_j}^\prime(r)Y_{l_j}^m(\theta,\phi), \quad
\rho^\prime=\sum_{j=1}^{j_{\rm max}}\rho_{l_j}^\prime(r)Y_{l_j}^m(\theta,\phi).
\end{eqnarray}
For stars having a purely poloidal magnetic field, the
oscillation modes are separated into even and odd modes. 
If the sets of the functions $(\xi_r, \xi_\phi,
B^\prime_\theta)$ and $(\xi_\theta,B^\prime_r,B^\prime_\phi)$ of a mode are respectively even
and odd (odd and even) functions about the equator, we call the mode an even (odd) mode.
In this paper, we usually use $j_{\rm max}=12$ to obtain solutions
with sufficiently high-angular resolution. Substituting the expansions
(11)-(17) into the linearized basic equations (\ref{eq:eqmot}) to (8), we obtain a
finite set of coupled linear ordinary differential equations for the
expansion coefficients $S_{l_j}(r)$, $T_{l_j^\prime}(r)$,
$b^S_{l_j^\prime}(r)$, $b^H_{l_j^\prime}(r)$, $b^T_{l_j^\prime}(r)$,
and $r\rd b^H_{l_j^\prime}(r)/\rd r$, which we call the oscillation
equations to be solved in the interior of magnetized stars. The
oscillation equations obtained for the star magnetized with a purely poloidal field are given in
Appendix A. 
The set of ordinary differential equations is solved
as an eigenvalue problem of $\sigma^2$ by applying boundary conditions
at the center and surface of the star (see also Appendix A).
Since the eigenvalue $\sigma^2$ is a real number for the boundary conditions we use,
the eigenvalue $\sigma^2>0$ corresponds to stable and purely oscillatory modes of the frequency $\pm\sigma$, and 
$\sigma^2<0$ to unstable and monotonically growing modes of the growth rate $\eta=\sqrt{-\sigma^2}$.

For a given $j_{\rm max}$, we find numerous solutions to the oscillation
equations.
Most of them, however, are dependent on $j_{\rm max}$. We have to
look for solutions that are independent of $j_{\rm max}$.

\section{Numerical results}

We use polytropes of index $n=1$ and 1.5 as a background model to
calculate non-axisymmetric ($m\neq 0$) oscillations of stars magnetized with a poloidal field. 
In this numerical study, magnetic modes, the eigenvalue $\sigma^2$ of which is proportional to $B_S^2$, 
are only the modes we discuss.
We cannot correctly compute $g$-, $f$-, and $p$-modes of the magnetized stars (see \S 6).
We find both stable (oscillatory) magnetic modes with $\sigma^2>0$ and unstable (monotonically growing)
magnetic modes with $\sigma^2<0$.
In fact, if we write $\sigma=\sigma_{\rm I}\rmi=\pm\eta \rmi$ with $\eta$ being a positive real number,
the time dependence of the modes is given by $\exp(\mp\eta t)$, and
the modes with $\exp(\eta t)$ monotonically grow with time without bound where $\eta$ may be regarded as the growth rate.
The modes with $\sigma^2<0$ may correspond to magnetic instability.
It is well known that the stars having purely poloidal magnetic
fields are unstable and the energy of the field is dissipated quickly,
that is, for several ten milliseconds (e.g., Markey \& Tayler 1973, van Assche, Goossens, Tayler 1982, Braithwaite 2007, 
Lasky et al. 2011; Ciolfi \& Rezzolla 2012).

\subsection{Stable Magnetic Modes}

In Figure 2, we plot the eigenfrequencies $\sigma$ of stable magnetic
modes that have no radial nodes of $S_{l_1}$ for $m=1$, 2, 3, and 4 versus the Alfv\'en frequency $\sigma_A$
defined as 
\begin{eqnarray}
\sigma_A\equiv B_{\rm S}/\sqrt{4\pi\rho_cR^2},
\end{eqnarray}
where $\rho_c$ is the central density, $R$ is the radius of the star,
and $B_{\rm S}=\mu_b/R^3$ is the magnetic field strength measured at the surface, and
the frequencies $\sigma$ and $\sigma_A$ are normalized 
by $\Omega_{\rm K}=\sqrt{GM/R^3}$ where $M$ is the mass of the star and $G$ is the gravitational constant.
In this paper, we assume $M=1.4M_\odot$ and $R=10^6$cm, for which 
we have the ratio $\sigma_A/\Omega_K=4.42\times 10^{-4}(B_S/10^{15}{\rm G})$.
In this paper we use the central density $\rho_c$ to define the Alfv\'en frequency $\sigma_A$.
We may use instead the mean density $\bar\rho\equiv M/(4\pi R^3/3)$ for the definition.
Since for polytropes
$
\bar\rho=-3\rho_c\zeta_1^{-1}\left(d{\Theta}/d{\zeta}\right)_{\zeta=\zeta_1}\equiv C\rho_c, 
$
where $\Theta(\zeta)$ is the Lane-Emden function and
$\Theta(\zeta_1)=0$, we have $C=3/\pi^2\approx 0.304$ and 0.167 for the indices $n=1$ and 1.5.
If we define $\sigma_A^*=B_S/\sqrt{4\pi\bar\rho R^2}$, we have
$\sigma_A^*=\sigma_A/\sqrt{C}$.
Note also that $\Omega_K=\sqrt{4\pi G\bar\rho/3}\propto\sqrt{G\bar\rho}$.
We find that the eigenfrequency of the
modes is isolated and proportional to the Alfv\'en frequency $\sigma_A$, that is, $B_S$. 
This property confirms that the modes we obtained are discrete magnetic modes.  
\begin{figure}
\begin{center}
\resizebox{0.5\columnwidth}{!}{
\includegraphics{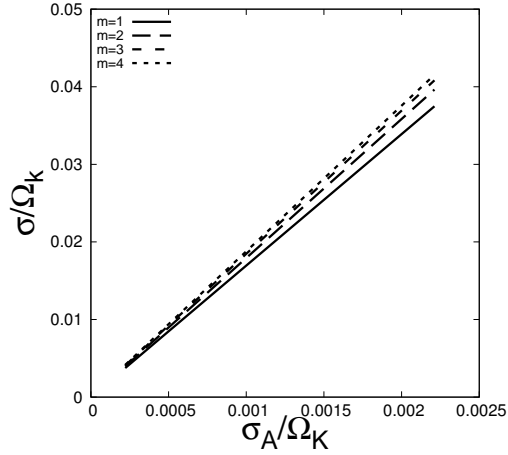}}
\end{center}
\caption{Eigenfrequency $\sigma$ of the stable magnetic modes of odd parity for $m=1$, $2$, $3$,
  and $4$ versus the Alfv\'en frequency
  $\sigma_A$ for the $n=1$ polytrope.}
\end{figure}
Note that we find stable magnetic modes only for odd parity and cannot find stable magnetic modes of even parity.

In Table 1, we tabulate the eigenfrequency $\bar\sigma\equiv \sigma/\Omega_K$
of stable magnetic modes for the polytrope of index $n=1$ and 1.5 for
$B_{\rm S}=10^{15}$ G.
Here, we have assumed that $\gamma=0$.
Note that the magnetic modes we can find for each value of $m$ are those that have only a few radial nodes of the expansion coefficient $S_l$, and it becomes difficult to find magnetic modes
as the number of radial nodes increases.
From Table 1, we find that for given $m$ and number of radial nodes, the normalized oscillation
frequency $\bar\sigma$ of the magnetic mode of $n=1.5$ is larger than that of $n=1$. 
For a given $m$, the oscillation frequency gradually decreases as the number of radial
nodes increases. This tendency of the oscillation
frequency is similar to that found by Asai \& Lee (2014) for axisymmetric ($m=0$) toroidal modes of the
stars magnetized with a purely poloidal magnetic field in general relativistic
framework. In addition, for a given number of radial nodes,
the larger the azimuthal wavenumber $m$, the larger the frequency
of the magnetic modes. 

\begin{table*}
\begin{center}
\caption{Normalized eigenfrequency $\bar{\sigma}$ of the stable magnetic modes of odd parity for $B_{\rm S}=10^{15}$ G.}
\begin{tabular}{@{}cccc}
\hline
 & & $n=1$ & \\
\hline
$m$ & & number of radial nodes & \\
\hline
 & 0 & 1 & 2 \\
$1$ & 0.007523 & 0.007120 & 0.006954\\
$2$ & 0.007943 & 0.007464 & 0.007216 \\
$3$ & 0.008174 & 0.007710 & 0.007439 \\
$4$ & 0.008310 & 0.007887 & 0.007618 \\
\hline
& & $n=1.5$ & \\
\hline
$m$ & & number of radial nodes & \\
\hline
& 0 & 1 & 2 \\
$1$ & 0.010304 & 0.009975 & 0.009817\\
$2$ & 0.010720 & 0.010247 & 0.010038 \\
$3$ & 0.010961 & 0.010485 & 0.010229\\
$4$ & 0.011106 & 0.010663 & 0.010397\\
\hline
\end{tabular}
\medskip
\end{center}
\end{table*}

In Figure 3, we plot the eigenfunctions of an $m=1$ magnetic mode that has no radial nodes of $S_{l_1}$
for the $n=1$ polytrope and $B_{\rm S}=10^{15}$ G. 
We find that the eigenfunctions
$\pmb{S}$, $\pmb{H}$, and $\pmb{T}$ of this mode have large
amplitudes in the the core. 
We also find that the horizontal $\pmb{H}$ and
toroidal $\pmb{T}$ components show rapid spacial oscillations near the stellar surface, 
although the radial component $\pmb{S}$
does not. This phenomena will be discussed in \S 5. 
Note that for isentropic ($\gamma=0$) models, we can obtain stable magnetic
modes even for magnetic fields as weak as $B_{\rm S}\sim 10^{12}$ G.

\begin{figure}
\begin{center}
\resizebox{0.39\columnwidth}{!}{
\includegraphics{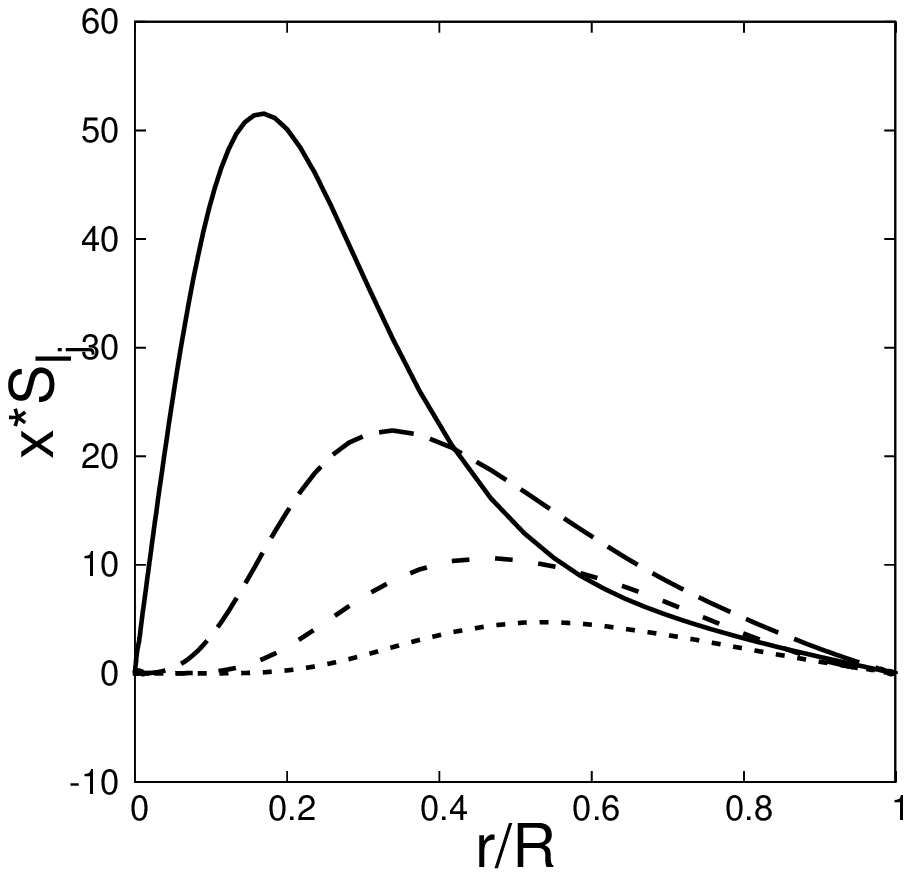}}
\hspace*{-1.75cm}
\resizebox{0.39\columnwidth}{!}{
\includegraphics{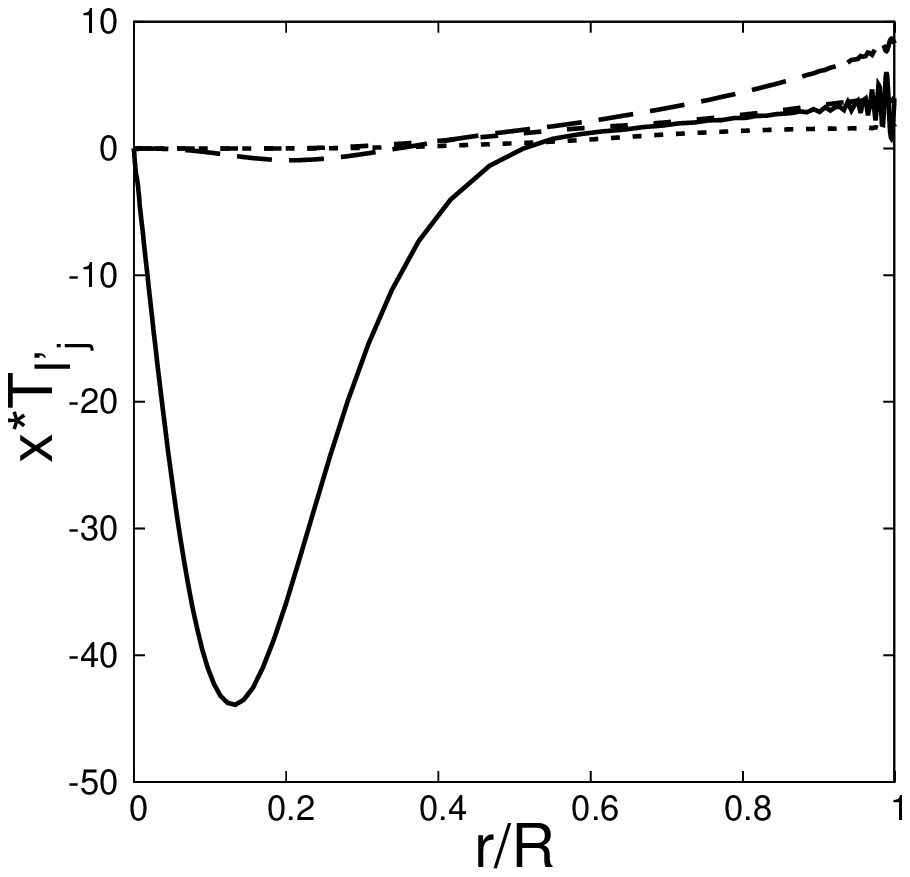}}
\hspace*{-1.75cm}
\resizebox{0.39\columnwidth}{!}{
\includegraphics{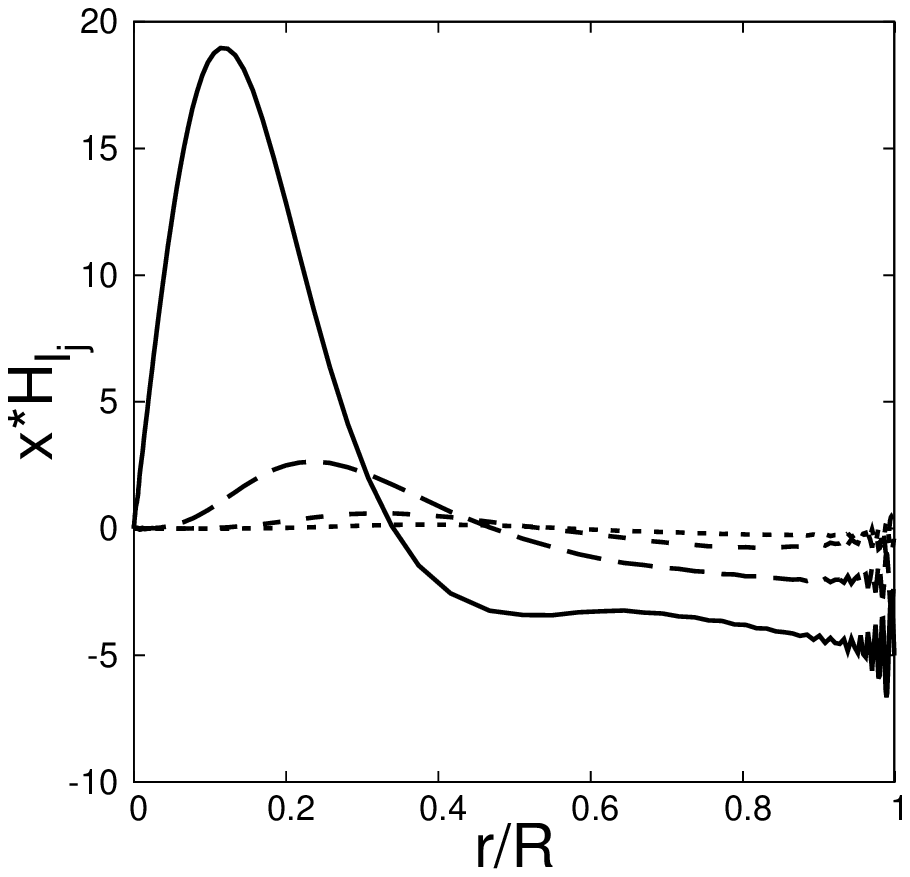}}
\end{center}
\caption{Expansion coefficients $xS_l$, $xT_{l^\prime}$, and $xH_l$ as
  a function of $x=r/R$ for an
  $m=1$ stable magnetic mode of odd parity for the polytrope with $n=1$ for
  $B_{\rm S}=10^{15}$
  G, where
  the solid lines, the long dashed lines, the short dashed lines, and
  the dotted lines are for the expansion coefficients associated with
  $l_j$ (or $l_j^\prime$) from $j=1$ to 4. 
  The amplitude normalization is given by $T_{l^\prime_{l_1}}=1$ at the surface.
  Here, the frequency
  $\bar\sigma\equiv\sigma/\Omega_{\rm K}$ of the mode is 0.007523.}
\end{figure}

Assuming $\phi=0$, we may define the spacial oscillation pattern $\hat{\pmb{\xi}}$ of the displacement vector $\pmb{\xi}$ as
\begin{eqnarray}
{\hat\xi}_j(r,\theta)=\xi_j(r,\theta,\phi=0),
\end{eqnarray}
where $j=r,~\theta,~\phi$, and
the patterns $\hat\xi_j(r,\theta)$ 
for an $m=1$ stable magnetic mode are shown in Figure 4, 
where the vertical $z$-axis given by $x=0$ is the symmetry axis,
and the amplitudes are normalized such that ${\rm max}(|\hat\xi_j(r,\theta)|)=1$ for $j=r,~\theta,~\phi$.
Since the mode is an odd mode, as shown by the figure, the 
patterns $\hat\xi_r(r,\theta)$ and $\hat{\xi}_\phi(r,\theta)$ are
antisymmetric about the equator given by $z=0$, while 
$\hat{\xi}_\theta(r,\theta)$ is symmetric.
The oscillation patterns have large amplitudes along the symmetry axis.
The $\phi$ component of $\hat{\pmb{\xi}}$ shows a pattern that reflects the existence of
magnetic fields closed within the interior.
\begin{figure}
\begin{center}
\resizebox{0.3\columnwidth}{!}{
\includegraphics[angle=0]{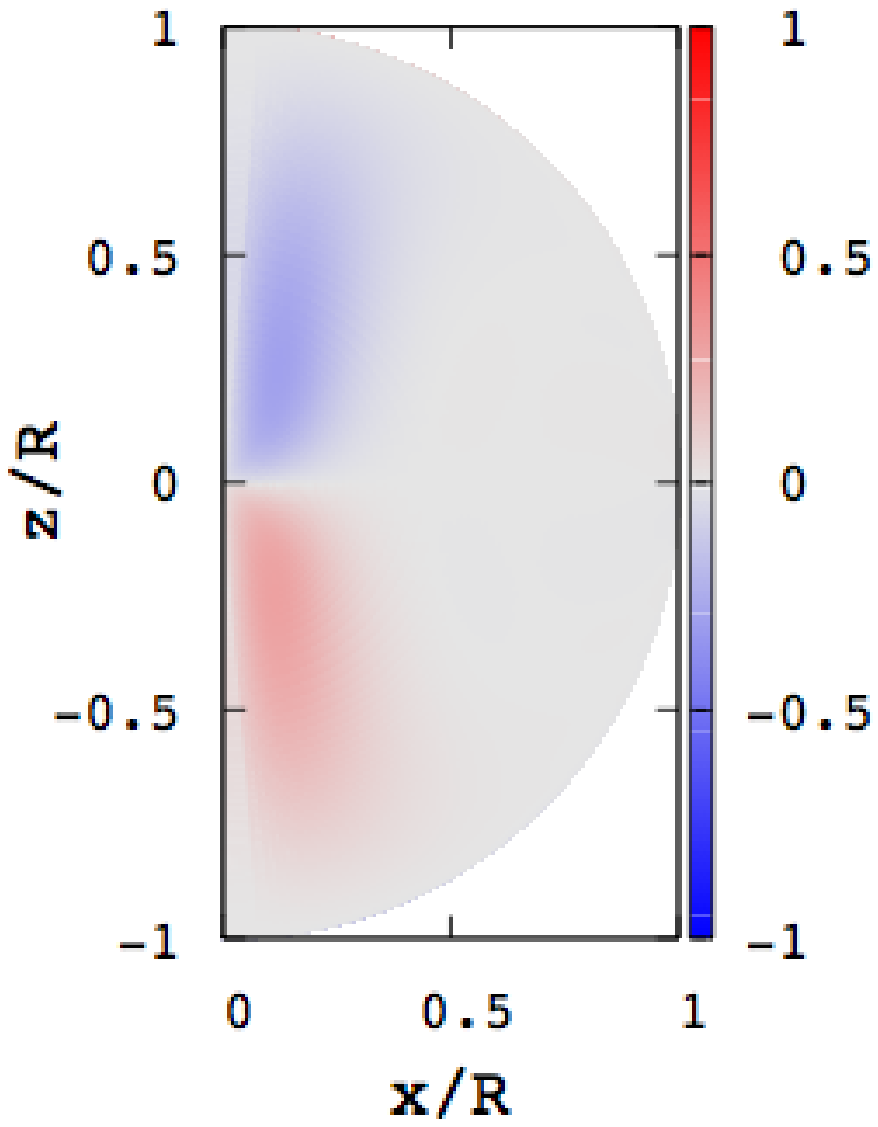}}
\hspace*{-0.35cm}
\resizebox{0.3\columnwidth}{!}{
\includegraphics[angle=0]{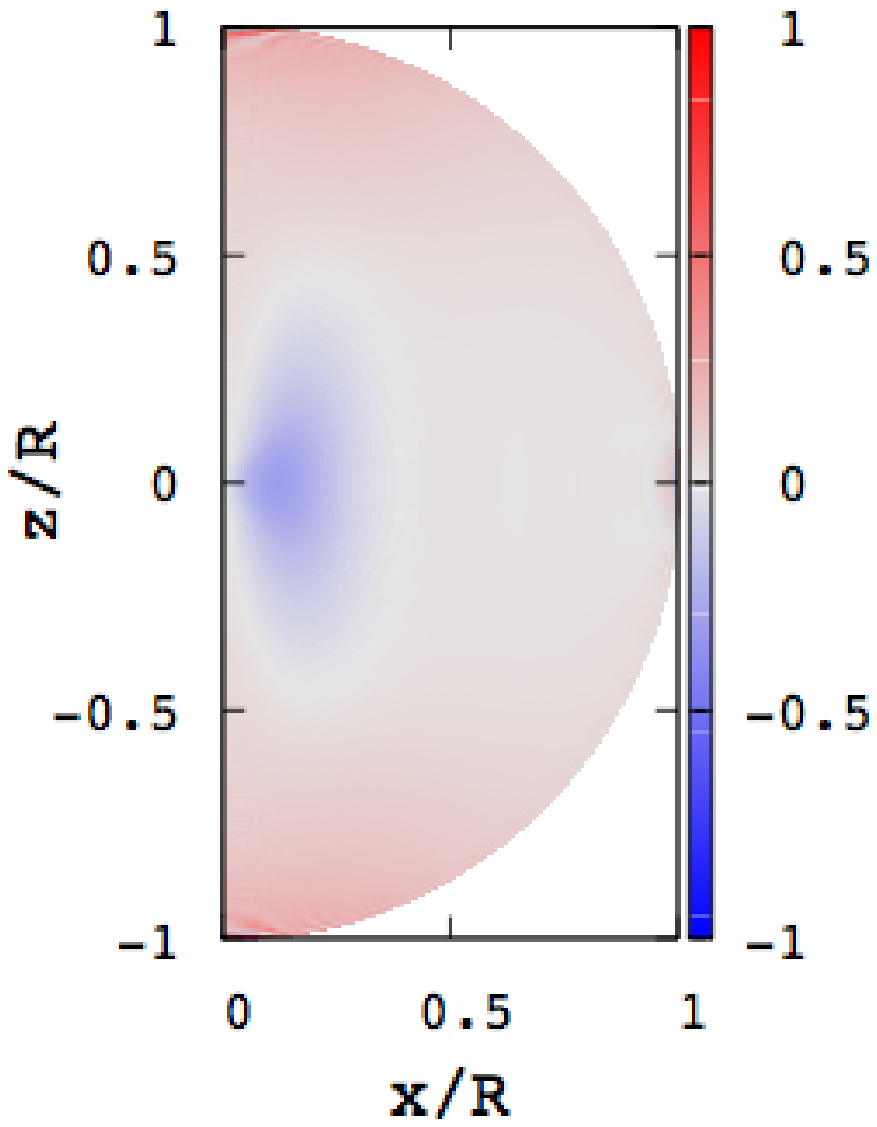}}
\hspace*{-0.35cm}
\resizebox{0.3\columnwidth}{!}{
\includegraphics[angle=0]{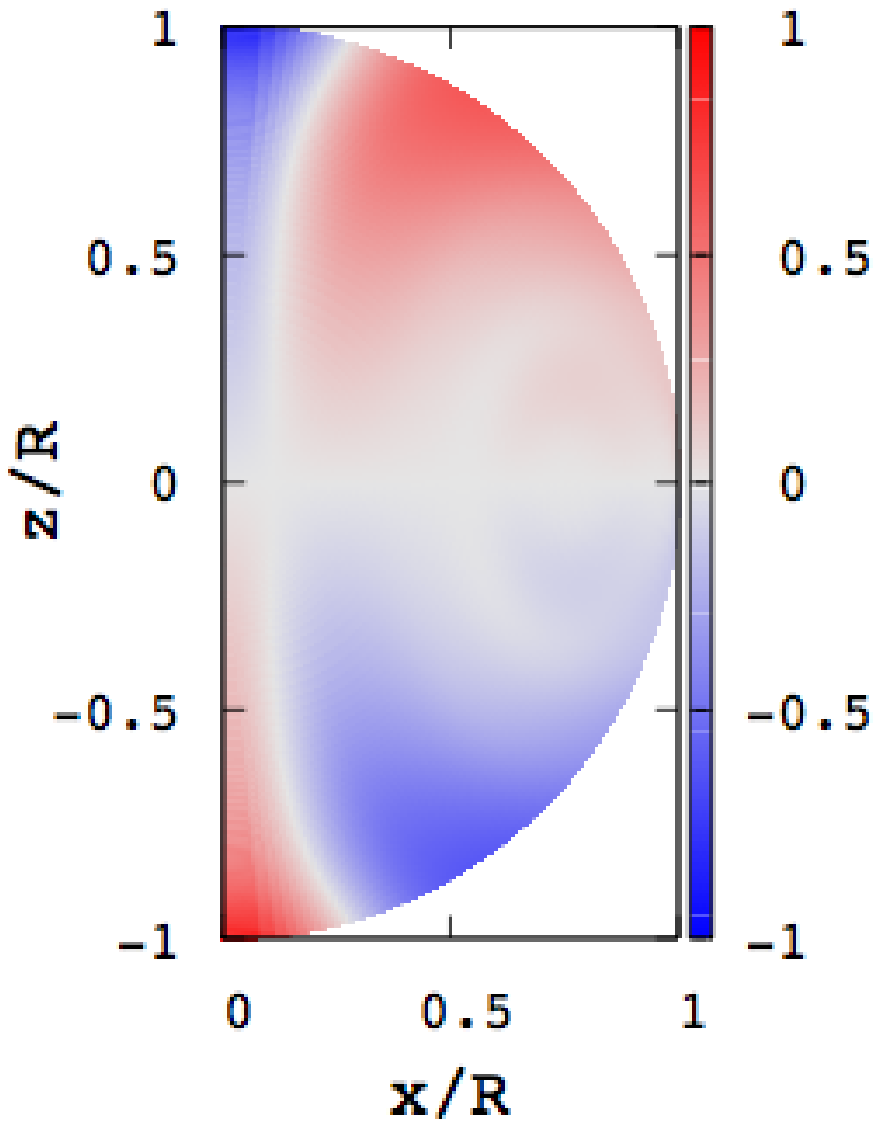}}
\caption{Spatial oscillation patterns $\hat\xi_r(r,\theta)$ (left),
  $\hat\xi_\theta(r,\theta)$ (middle), and
  $\hat\xi_\phi(r,\theta)$ (right)
  of the $m=1$ stable magnetic mode of Figure 3, where the amplitudes are normalized such that
${\rm max}(|\hat\xi_j(r,\theta)|)=1$.}
\end{center}
\end{figure}

In Table 2, we tabulate the ratios 
$|\hat{\xi}_\theta^{\rm max}|/|\hat\xi_r^{\rm max}|$ and $|\hat{\xi}_\phi^{\rm max}|/|\hat\xi_r^{\rm max}|$ for
the nodeless magnetic modes of the $n=1$
and $n=1.5$ polytropes, where $|\hat\xi_j^{\rm max}|={\rm max}(|\hat\xi_j(r,\theta)|)$, and we have assumed $\gamma=0$ and $B_{\rm S}=10^{15}$ G.
From Table 2, we find that the relative amplitudes $|\hat{\xi}_\theta^{\rm
  max}|/|\hat\xi_r^{\rm max}|$ and $|\hat{\xi}_\phi^{\rm
  max}|/|\hat{\xi}_r^{\rm max}|$
are rather insensitive to the azimuthal wavenumber $m$, and that the components $\hat\xi_r$ and $\hat\xi_\theta$
dominate $\hat{\xi}_\phi$.
The pressure perturbation $\delta U\equiv p^\prime(r,\theta,\phi=0)/\rho gr$ has almost negligible amplitudes compared to 
the displacement vector. 
The oscillation patterns of the nodeless mode are almost the same for the $n=1$ and $n=1.5$
polytropes although the ratios $|\hat{\xi}_\theta^{\rm max}|/|\hat\xi_r^{\rm max}|$ and 
$|\hat{\xi}_\phi^{\rm max}|/|\hat\xi_r^{\rm max}|$ for $n=1$ are slightly larger than those for $n=1.5$ as indicated by Table 2. 
For the magnetic modes that have non-zero radial nodes of $S_{l_1}$, 
the tendencies of the amplitude ratios are found similar to
those for the nodeless magnetic modes, and 
the oscillation patterns 
look quite similar for the polytropes of $n=1$ and 1.5

As $|m|$ increases, the oscillation amplitudes tend to be confined in the envelope region away from the symmetry axis,
as exemplified by the spatial oscillation patterns of an $m=5$ unstable magnetic mode shown in Figure 8.

\begin{table*}
\begin{center}
\caption{Amplitude ratios between the components of the displacement vector of stable
  magnetic modes of odd parity for $B_{\rm S}=10^{15}$ G}
\begin{tabular}{@{}cccc}
\hline
 & & $n=1$ & \\
\hline
 $m$ & $|\hat{\xi}_\theta^{\rm
   max}|/|\hat\xi_r^{\rm max}|$ & $|\hat{\xi}_\phi^{\rm max}|/|\hat\xi_r^{\rm
   max}|$ & $|\delta U^{\rm max}|/|\hat\xi_r^{\rm max}|$\\
\hline
$1$ & 1.109 & 0.322 & 9.833$\times 10^{-6}$\\
$2$ & 0.787 & 0.304 & 1.236$\times 10^{-5}$\\
$3$ & 1.234 & 0.366 & 1.408$\times 10^{-5}$\\
$4$ & 1.121 & 0.396 & 1.595$\times 10^{-5}$\\
\hline
& & $n=1.5$ & \\
\hline
$1$ & 2.552 & 0.111 & 6.954$\times 10^{-6}$\\
$2$ & 0.897 & 0.047 & 3.358$\times 10^{-6}$\\
$3$ & 1.124 & 0.051 & 5.376$\times 10^{-6}$\\
$4$ & 1.365 & 0.103 & 7.191$\times 10^{-6}$\\
\hline
\end{tabular}
\medskip
\end{center}
\end{table*}

\subsection{Unstable Magnetic Modes}

\begin{table*}
\begin{center}
\caption{Growth rate $\bar\eta\equiv\eta/\Omega_{\rm{K}}$ of the
  monotonically growing magnetic modes of even and odd parities for
  $B_{\rm{S}}=10^{15}$ G}
\begin{tabular}{@{}cccc|cccc}
\hline
 & & & & $n=1$ & & & \\
\hline
 & & even parity & & & & odd parity & \\
\hline
$m$ & & number of radial nodes & & $m$ & & number of radial nodes & \\
\hline
 & 1 & 2 & 3 & & 1 & 2 & 3 \\
1 & 0.000961 & 0.000580 & 0.000405 & 1 &  & & \\
2 & 0.001557 & 0.000977 & 0.000709 & 2 & 0.003562 & & \\
3 & 0.001964 & 0.001273 & 0.000949 & 3 & 0.004435 & 0.002038 & \\
4 & 0.002260 & 0.001505 & 0.001144 & 4 & 0.004879 & 0.003088 & \\
5 & 0.002488 & 0.001695 & 0.001452 & 5 & 0.005147 & 0.003685 &
0.002008 \\
6 & 0.002676 & 0.001859 & 0.001452 & 6 & 0.005326 & 0.004086 &
0.002728 \\
7 & 0.002852 & 0.002110 & 0.001889 & 7 & 0.005454 & 0.004377 &
0.003209 \\
8 & 0.003096 & 0.002616 & 0.002069 & 8 & 0.005548 & 0.004598 &
0.003566 \\
\hline
 & & & & $n=1.5$ & & & \\
\hline
 & & even parity & & & & odd parity & \\
\hline
$m$ & & number of radial nodes & & $m$ & & number of radial nodes & \\
\hline
 & 1 & 2 & 3 & & 1 & 2 & 3 \\
1 & 0.000898 & 0.000430 & & 1 & & & \\
2 & 0.001601 & 0.000804 & & 2 & & & \\
3 & 0.002114 & 0.001121 & & 3 & & & \\
4 & 0.002482 & 0.001386 & & 4 & 0.006170 & 0.003683 & \\
5 & 0.002769 & 0.001618 & & 5 & 0.006510 & 0.004506 & 0.002007 \\
6 & 0.003328 & 0.002611 & & 6 & 0.006734 & 0.005047 & 0.003147 \\
7 & 0.004089 & 0.002895 & & 7 & 0.006892 & 0.005435 & 0.003849 \\
8 & 0.004659 & 0.003053 & & 8 & 0.007009 & 0.005727 & 0.004354 \\
\hline
\end{tabular}
\medskip
\end{center}
\end{table*}

Unstable magnetic modes are found both for even parity and for odd parity. 
If we define the growth rate
$\eta>0$ such that $\sigma=\pm\eta \rmi$,
$\eta$
is almost exactly proportional to the field strength $B_{\rm S}$, particularly for the
case of isentropic $(\gamma=0)$ models.
In Table 3, we tabulate the normalized growth rate $\bar\eta\equiv \eta/\Omega_{\rm K}$
of unstable magnetic modes for the polytropes of index $n=1$ and 1.5, where
we have assumed $\gamma=0$ and $B_{\rm S}=10^{15}$ G.
Note that unstable magnetic modes are found even for magnetic fields
as weak as $B_{\rm S}\sim 10^{12}$ G. 
From Table 3, we find that the growth rates $\bar\eta$ of the
unstable magnetic modes for the $n=1$ and $n=1.5$ polytropes are quite similar, and that
it becomes difficult to obtain unstable magnetic modes as the number of radial nodes increases, 
particularly for $n=1.5$. 
Between $m=1$ and $m=8$,
we find unstable magnetic modes of odd parity only for $m\ge2$ for $n=1$
and for $m\ge 4$ for $n=1.5$. 
For a given $m$, the growth rate $\bar\eta$ gradually
decreases as the number of radial nodes of $S_{l_1}$ increases.
On the other hand,
for a given number of radial nodes, the larger the azimuthal index $m$, the
larger the growth rate of the unstable modes. 
Using the normalized growth rate $\bar\eta$, the growth time scale $\tau_g=1/\eta$ may be given by
\be
\tau_g=7.33\times 10^{-5}/\bar\eta \quad {\rm sec},
\ee
where we have assumed $M=1.4M_\odot$ and $R=10^6$cm.
For a typical value of $\bar\eta\sim 10^{-3}$, the growth time scale may be
$\tau_g\sim 5\times 10^{-2}$sec, which is consistent with the results, e.g., by Lasky et al. (2011) and
Ciolfi \& Rezzolla (2012).

Figure 5 shows the eigenfunctions of an $m=1$ unstable magnetic
mode of even parity that has one radial node of $S_{l_1}$ for the $n=1$
polytrope for $B_{\rm S}=10^{15}$ G, where the amplitudes are normalized by
$T_{l^\prime_1}=1$ at the surface.
We find that $S_{l_1}$ has large amplitudes in the core, while $T_{l^\prime_1}$ in the envelope.
The function $H_{l_1}$ has large
amplitudes both in the core and in the envelope of the star.
Note that the first components of $\pmb{S}$, $\pmb{T}$, and $\pmb{H}$ are dominating 
the other components.
\begin{figure}
\begin{center}
\resizebox{0.39\columnwidth}{!}{
\includegraphics{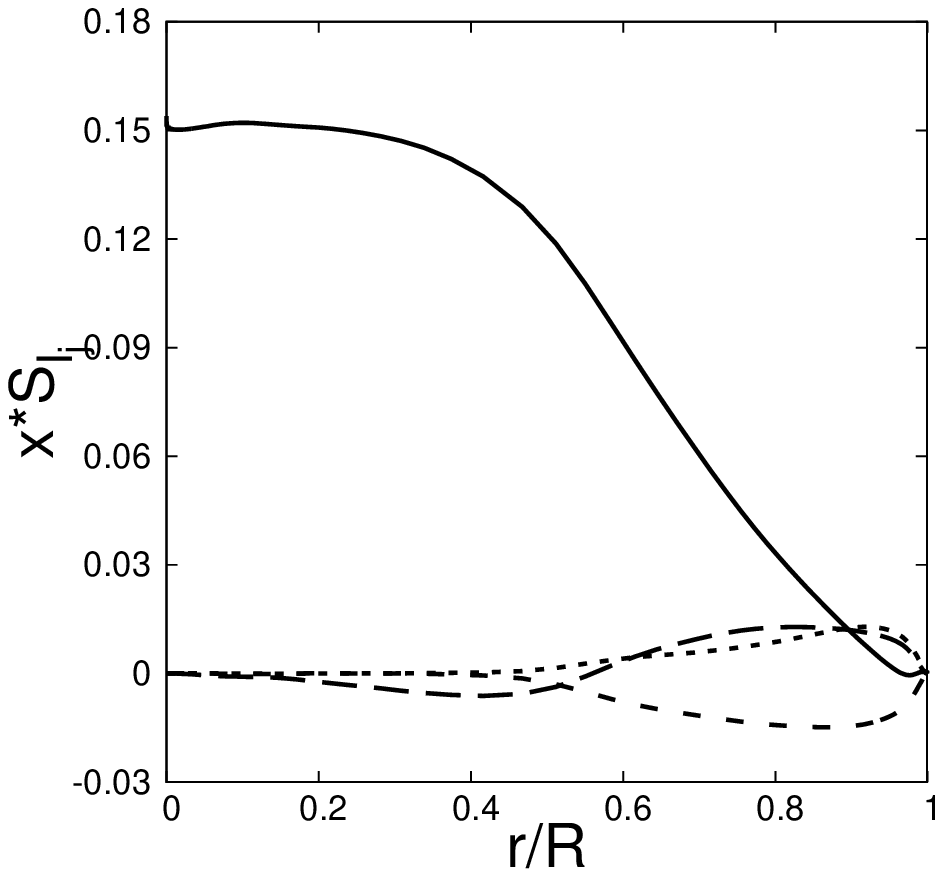}}
\hspace*{-1.75cm}
\resizebox{0.39\columnwidth}{!}{
\includegraphics{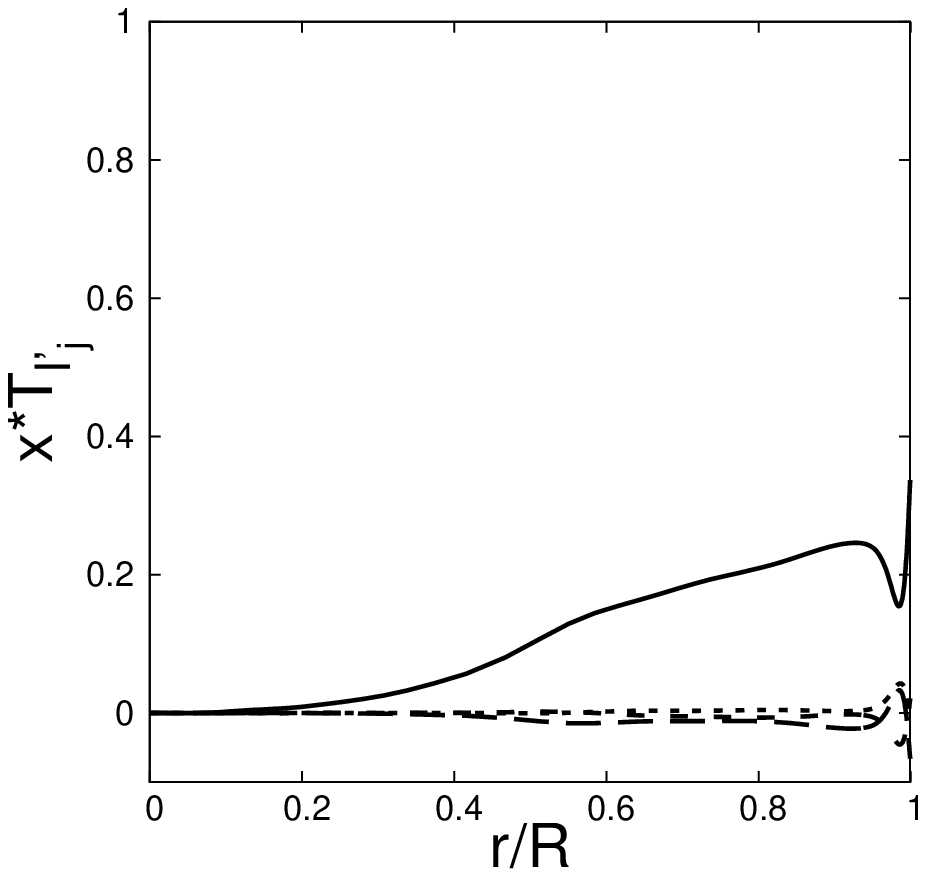}}
\hspace*{-1.75cm}
\resizebox{0.39\columnwidth}{!}{
\includegraphics{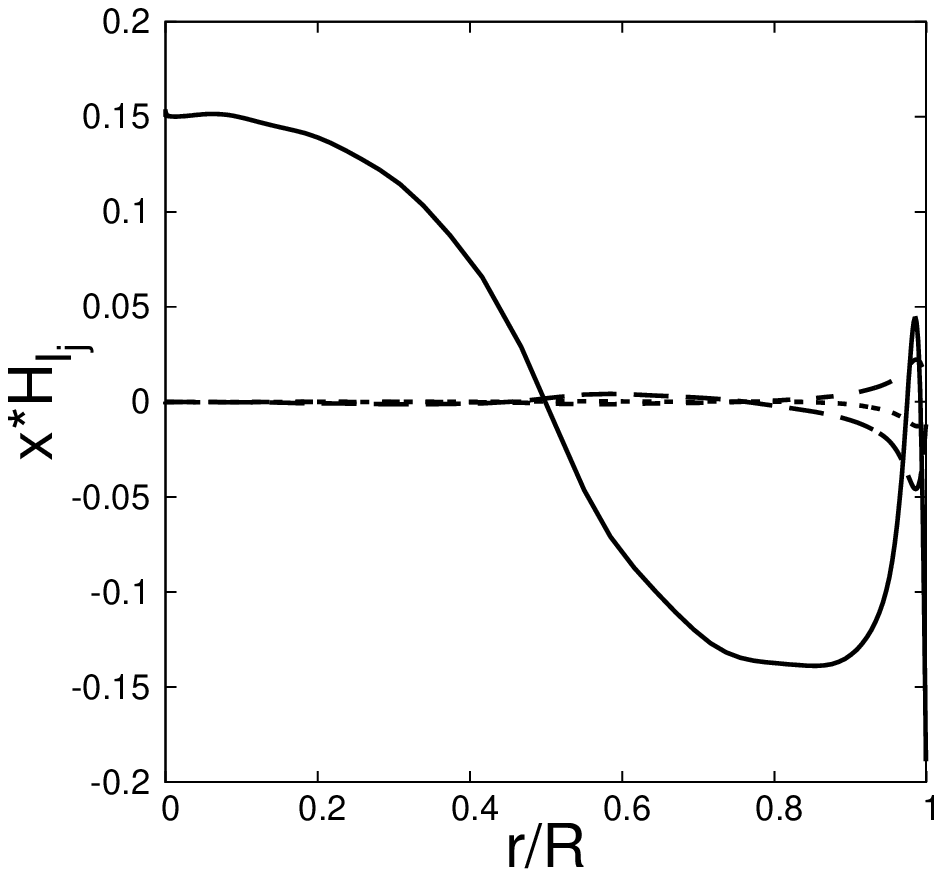}}
\end{center}
\caption{Same as Figure 3 but for an
  $m=1$ unstable magnetic mode of even parity with the growth rate $\bar\eta=0.00096$.}
\end{figure}

\begin{figure}
\begin{center}
\resizebox{0.32\columnwidth}{!}{
\includegraphics[angle=0]{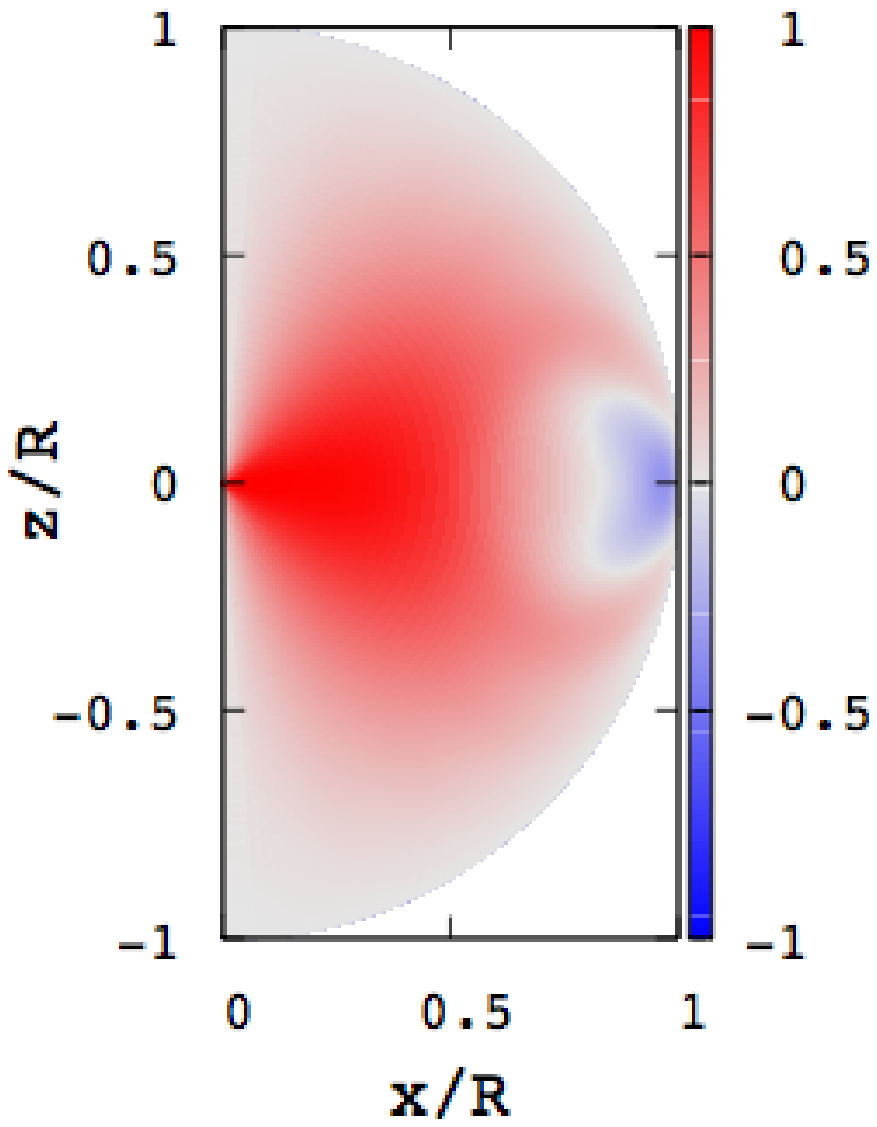}}
\hspace*{-0.44cm}
\resizebox{0.32\columnwidth}{!}{
\includegraphics[angle=0]{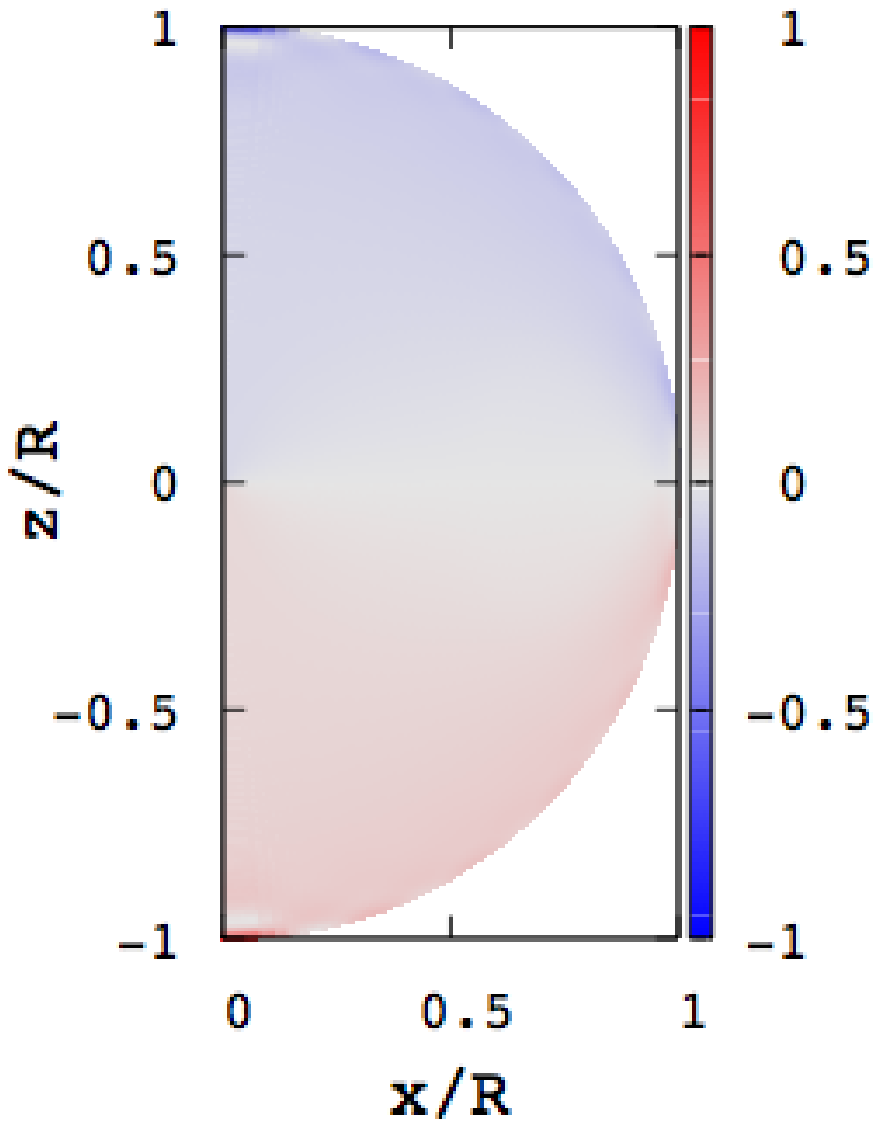}}
\hspace*{-0.4cm}
\resizebox{0.3\columnwidth}{!}{
\includegraphics[angle=0]{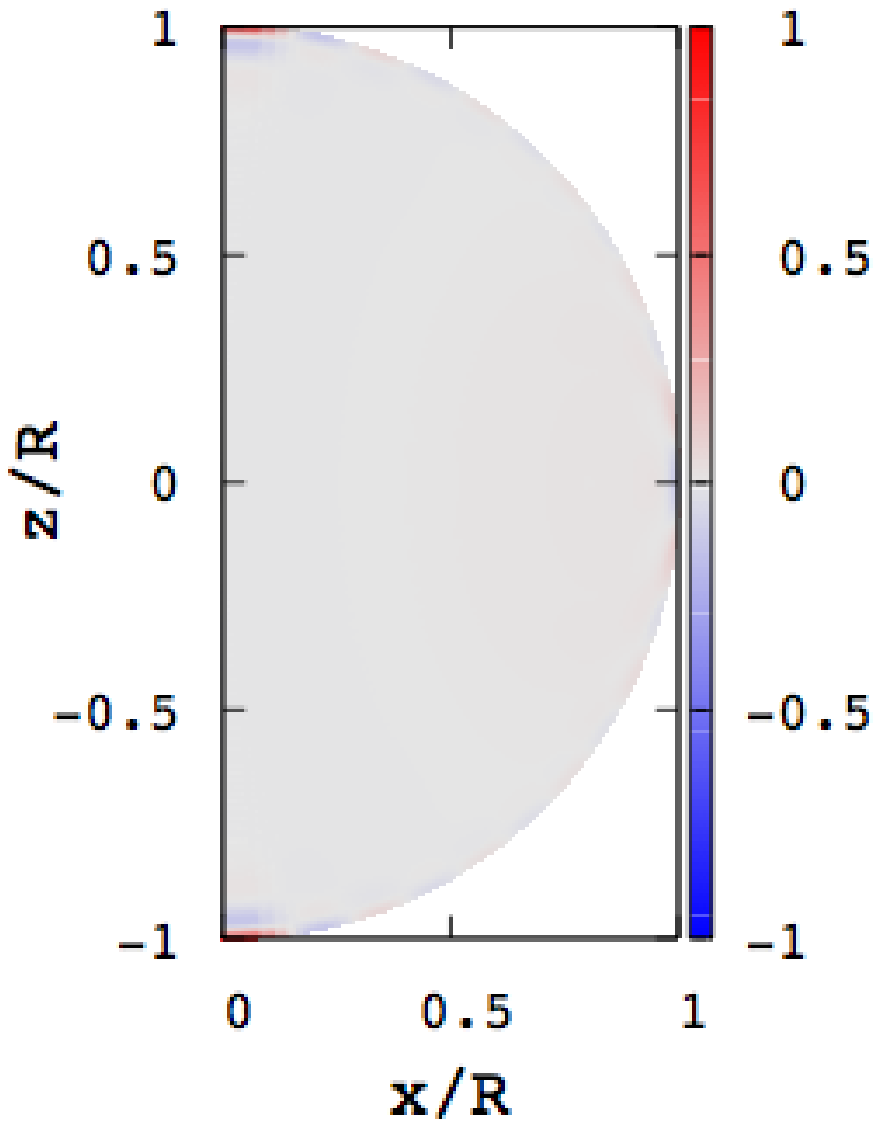}}
\caption{Same as Figure 4 but for the $m=1$ unstable magnetic mode of Figure 5.}
\end{center}
\end{figure}

The oscillation patterns $\hat{\xi}_r(r,\theta)$, $\hat{\xi}_\theta
(r,\theta)$, and $\hat{\xi}_\phi(r,\theta)$ for the $m=1$ unstable magnetic mode are given in Figure 6.
Since this mode has even parity, $\hat{\xi}_r$ and 
$\hat{\xi}_\phi$ are symmetric about the equator, while $\hat{\xi}_\theta$ is antisymmetric. 
The patterns $\hat{\xi}_r$ has large amplitudes along the equator, but
the amplitudes of 
the $\theta$ and $\phi$ components of $\hat{\pmb{\xi}}$ are confined to
the polar regions.
The existence of closed magnetic fields is recognized in the oscillation pattern of $\hat\xi_r$.

In Table 4, the ratios $|\hat{\xi}_\theta^{\rm
  max}|/|\hat{\xi}_r^{\rm max}|$ and $|\hat{\xi}_\phi^{\rm
  max}|/|\hat{\xi}_r^{\rm max}|$ are tabulated for the unstable magnetic modes
of the $n=1$ and $n=1.5$ polytropes for $B_{\rm S}=10^{15}$ G. From Table
4, we find that relative amplitudes $|\hat{\xi}_\theta^{\rm
  max}|/|\hat{\xi}_r^{\rm max}|$ and $|\hat{\xi}_\phi^{\rm
  max}|/|\hat{\xi}_r^{\rm max}|$ decrease with the azimuthal
wavenumber $m$, and that $\hat{\xi}_\phi$ dominates the other components.
The pressure perturbation $\delta U$ have 
negligible amplitudes compared to the displacement vector as in the case of
stable magnetic modes. 
The oscillation patterns are almost the same for the
$n=1$ and $n=1.5$ polytropes.

\begin{table*}
\begin{center}
\caption{Amplitude ratios between the components of the displacement vector of
  the unstable magnetic modes of even parity for $B_{\rm S}=10^{15}$ G}
\begin{tabular}{@{}cccc}
\hline
 & & $n=1$ & \\
\hline
 $m$ & $|\hat{\xi}_\theta^{\rm
   max}|/|\hat\xi_r^{\rm max}|$ & $|\hat{\xi}_\phi^{\rm max}|/|\hat\xi_r^{\rm
   max}|$ & $|\delta U^{\rm max}|/|\hat\xi_r^{\rm max}|$\\
\hline
$1$ & 17.05 & 319.1 & 3.462$\times 10^{-4}$\\
$2$ & 10.95 & 120.5 & 3.706$\times 10^{-4}$\\
$3$ & 9.183 & 72.64 & 3.137$\times 10^{-4}$\\
$4$ & 7.996 & 49.24 & 2.538$\times 10^{-4}$\\
\hline
& & $n=1.5$ & \\
\hline
$1$ & 31.98 & 685.1 & 1.471$\times 10^{-4}$\\
$2$ & 14.39 & 159.9 & 1.166$\times 10^{-4}$\\
$3$ & 7.662 & 63.72 & 7.266$\times 10^{-5}$\\
$4$ & 10.66 & 69.91 & 1.082$\times 10^{-4}$\\
\hline
\end{tabular}
\medskip
\end{center}
\end{table*}

Figures 7 and 8 show the eigenfunctions and spatial oscillation patterns of an $m=5$ unstable magnetic
mode of odd parity that has one radial node of $S_{l_1}$ for the $n=1$
polytrope for $B_{\rm S}=10^{15}$ G. 
The toroidal components $T_{l_j}$ of the displacement vector $\pmb{\xi}$ look similar to
those of the $m=1$ unstable mode of even parity in Figure 5, but the amplitudes of $S_{l_j}$ and $H_{l_j}$ are
confined in the envelope region, in contast to those of the $m=1$ magnetic mode.
This amplitude confinement to the envelope region is clearly seen in the spatial oscillation patterns shown
in Figure 8.
It is also interesting to note that the mode has negligible amplitudes along the symmetry axis.
\begin{figure}
\begin{center}
\resizebox{0.39\columnwidth}{!}{
\includegraphics{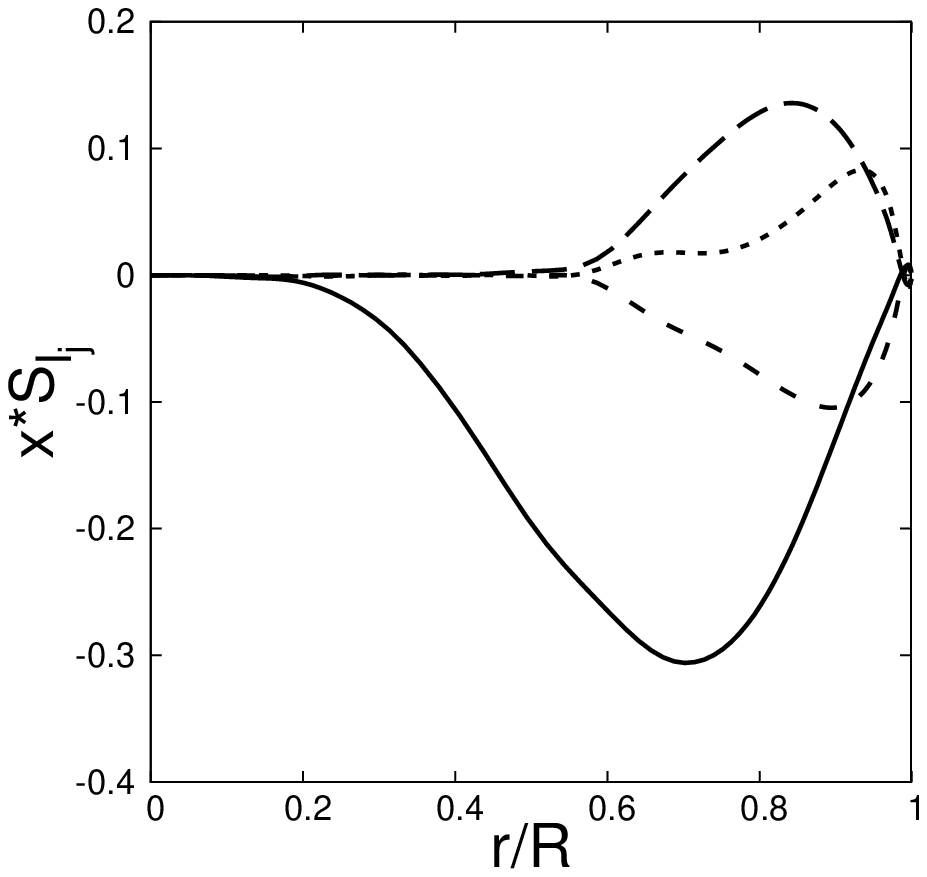}}
\hspace*{-1.75cm}
\resizebox{0.39\columnwidth}{!}{
\includegraphics{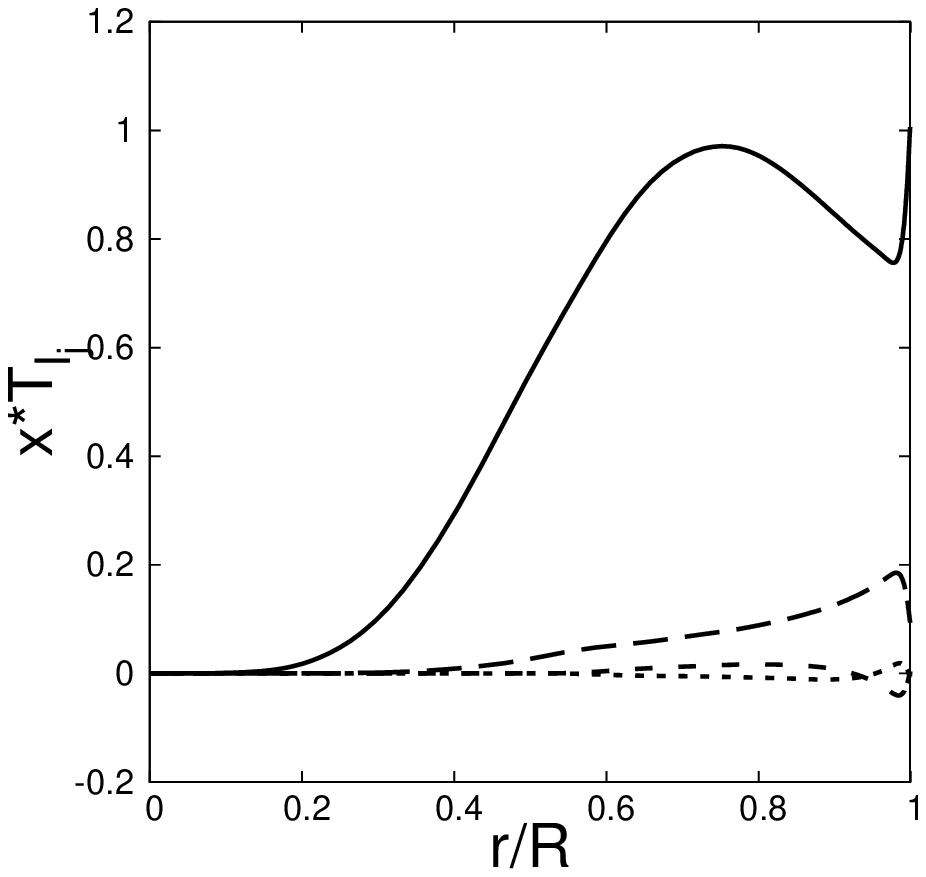}}
\hspace*{-1.75cm}
\resizebox{0.39\columnwidth}{!}{
\includegraphics{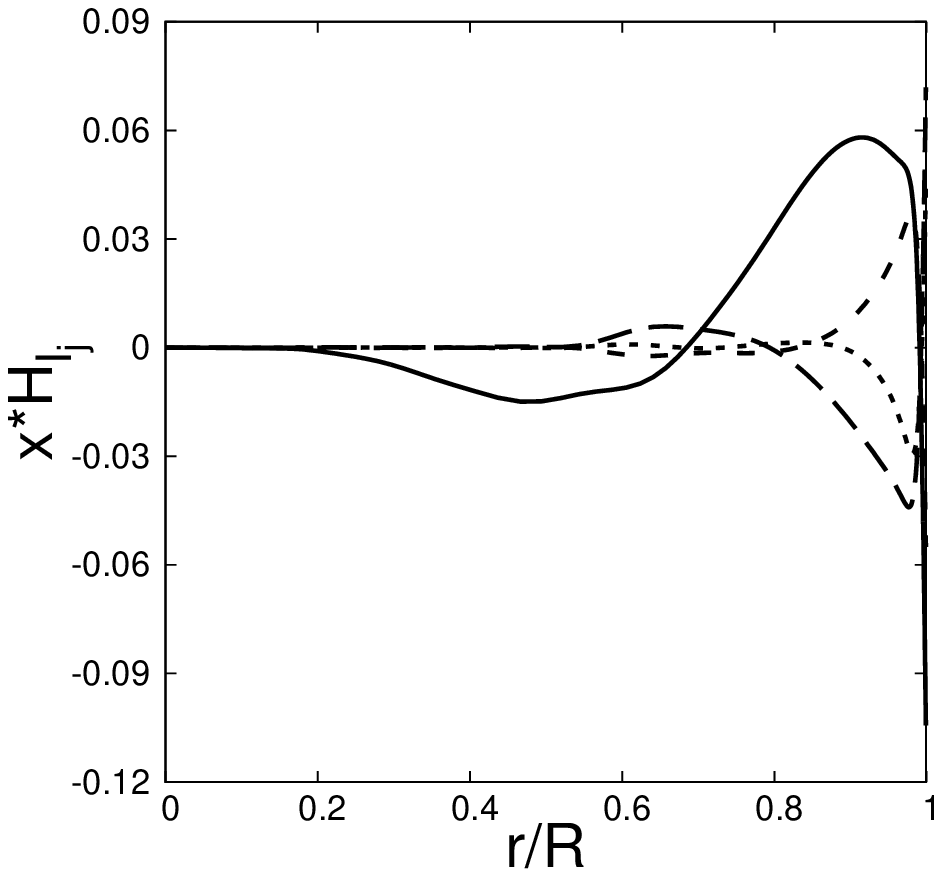}}
\end{center}
\caption{Same as Figure 3 but for 
an $m=5$ oscillatory magnetic modes of odd parity with the growth rate $\bar\eta=0.005147$.}
\end{figure}

\begin{figure}
\begin{center}
\resizebox{0.33\columnwidth}{!}{
\includegraphics[angle=0]{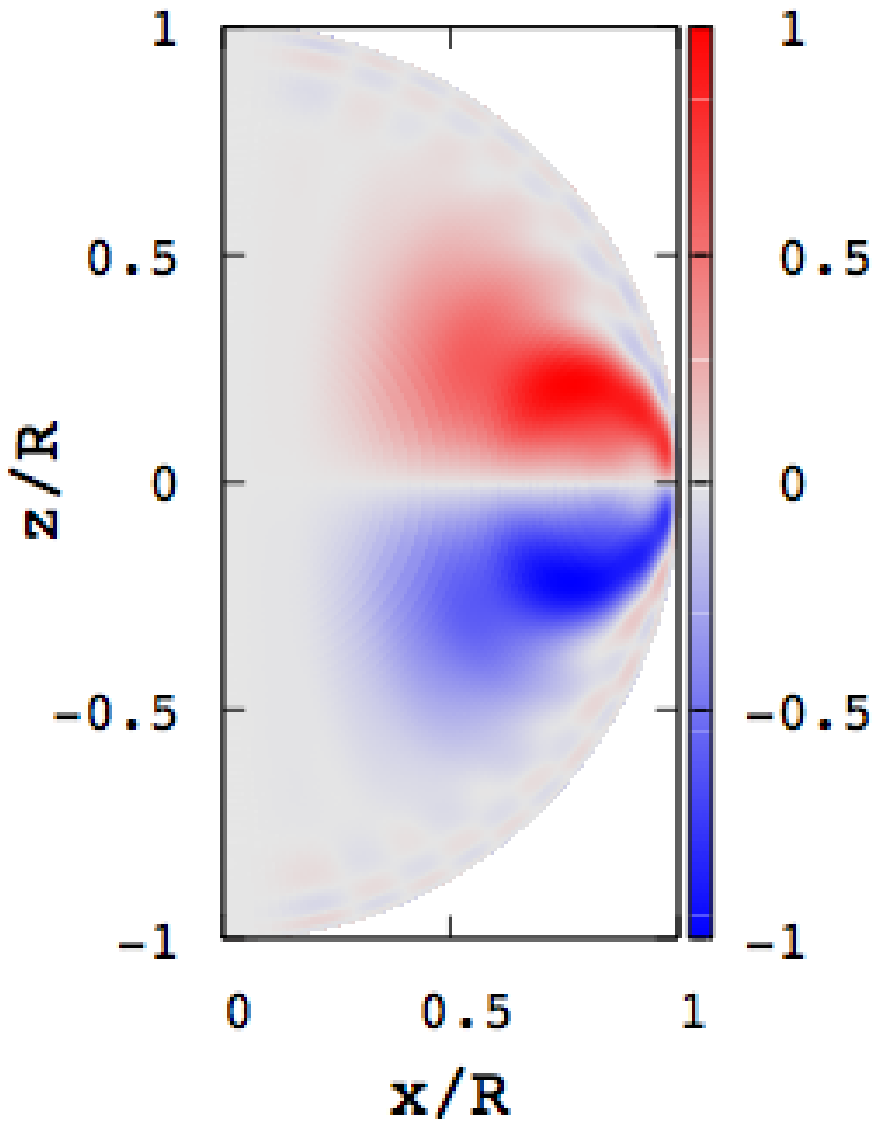}}
\hspace*{-0.44cm}
\resizebox{0.3\columnwidth}{!}{
\includegraphics[angle=0]{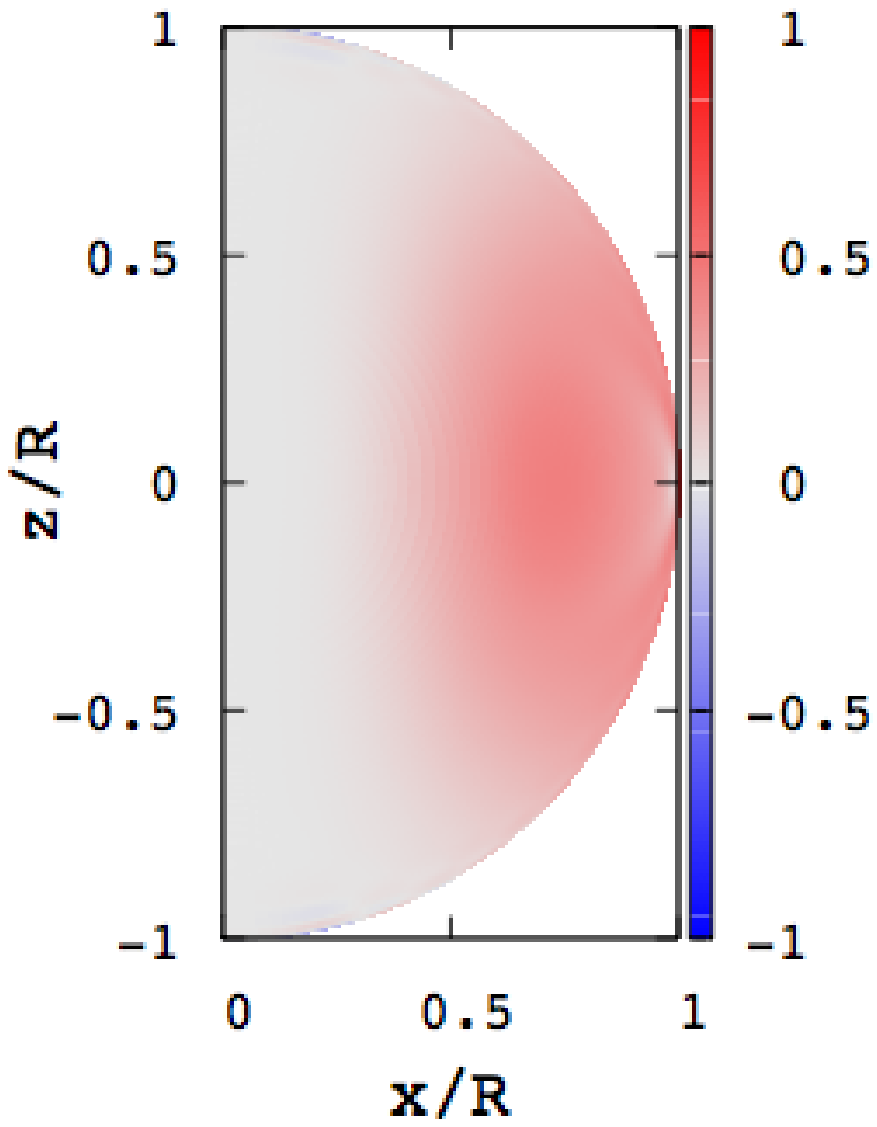}}
\hspace*{-0.4cm}
\resizebox{0.3\columnwidth}{!}{
\includegraphics[angle=0]{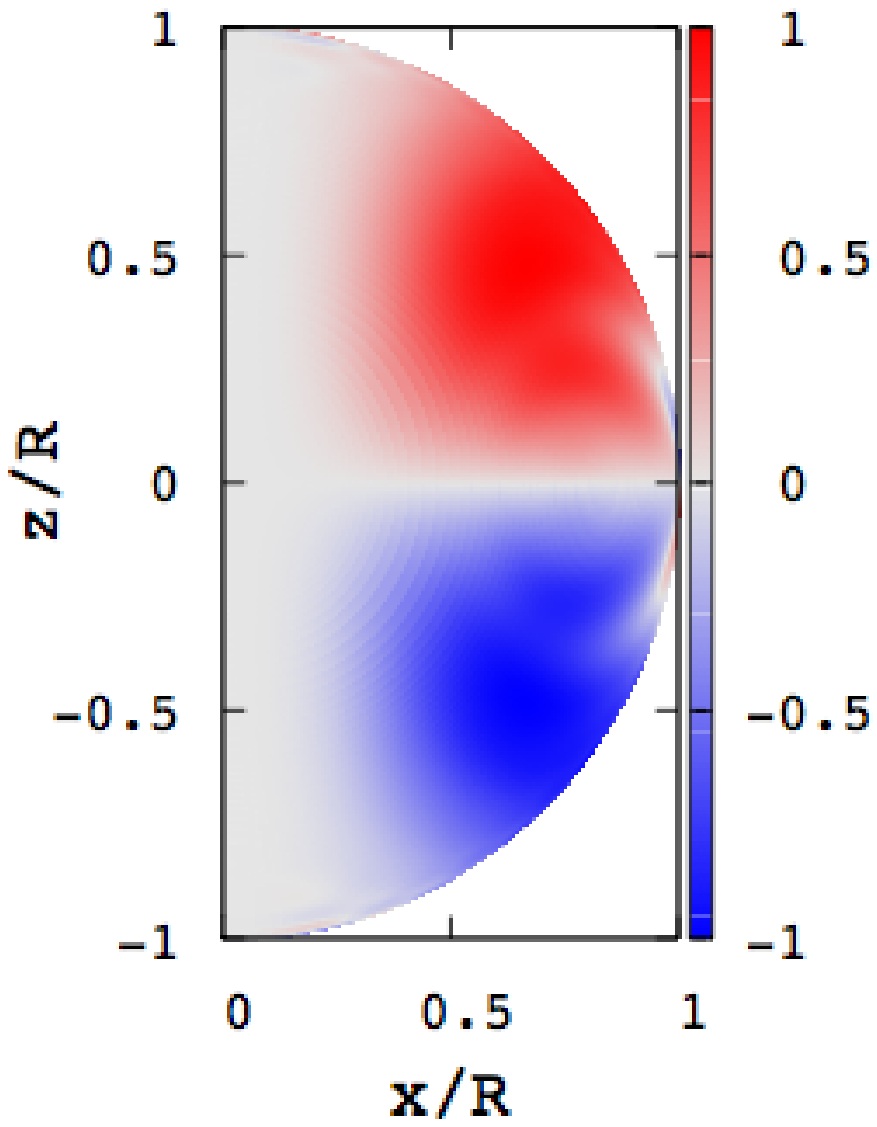}}
\caption{Same as Figure 4 but for 
the $m=5$ unstable magnetic mode of Figure 7.}
\end{center}
\end{figure}

\section{Discussion}

Magnetic modes depend on $\gamma$ as shown by Figure 9, in which the frequency $\sigma$ of 
a stable $m=2$ magnetic mode of odd parity with no radial nodes of $S_{l_1}$ and
the growth rate $\eta/\sigma_A$ of an unstable $m=2$ magnetic mode of even parity are plotted as a function of 
$\sqrt{|\gamma|}$ for the $n=1$ polytrope and for $B_{\rm S}=10^{15}$ G.
As $|\gamma|$ increases, the frequency $\sigma$ gradually increases (decreases) for $\gamma<0$ (for $\gamma>0$),
and we find no magnetic mode for $\sqrt{|\gamma|}/\sigma_A\ga 10$.
On the other hand, the growth rate $\eta$ decreases (increases) as
$|\gamma|$ increases for $\gamma<0$ ($\gamma>0$), and we find no unstable solutions beyond $\sqrt{|\gamma|}/\sigma_A\ga 4$.
For radiative stars with $\gamma<0$, the buoyant force tends to stabilize the magnetic instability.
It is important to note that when we normalize the eigenvalue $\sigma$ (or $\eta$) and 
the Brunt-V\"ais\"al\"a frequency $N\propto\sqrt{|\gamma|}$
in terms of $\sigma_A$,
the relation between $\sigma/\sigma_A$ (or $\eta/\sigma_A$) and 
$|\gamma|^{1/2}/\sigma_A$ in Figure 9 is almost independent of the magnetic field strength $B_S$.

\begin{figure}
\begin{center}
\resizebox{0.45\columnwidth}{!}{
\includegraphics{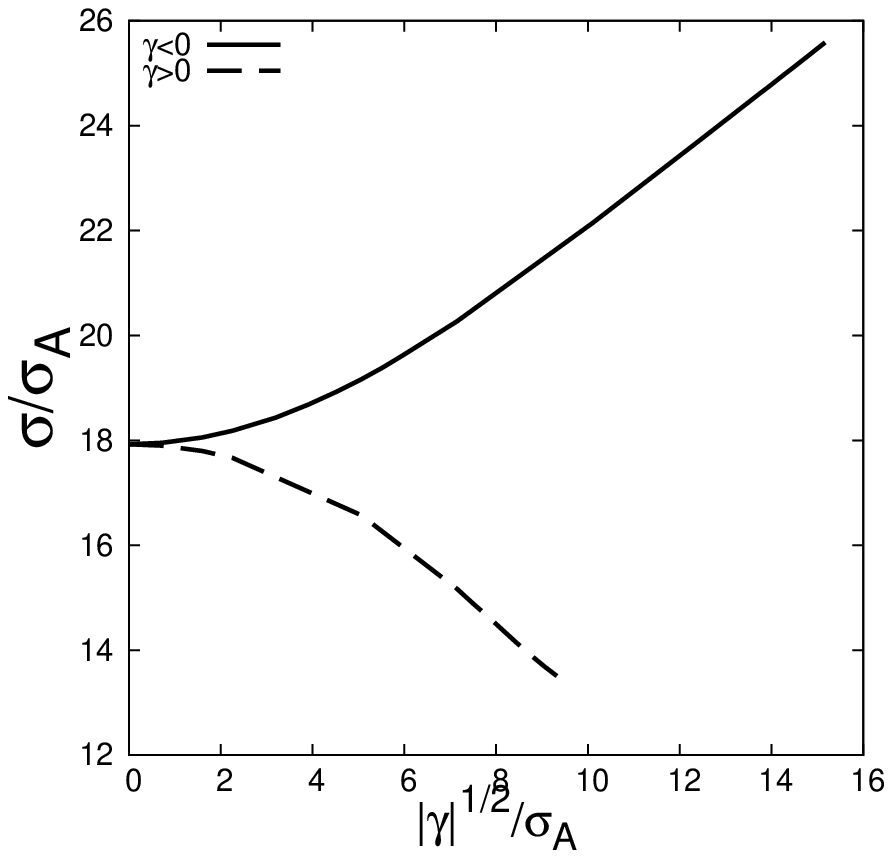}}
\resizebox{0.45\columnwidth}{!}{
\includegraphics{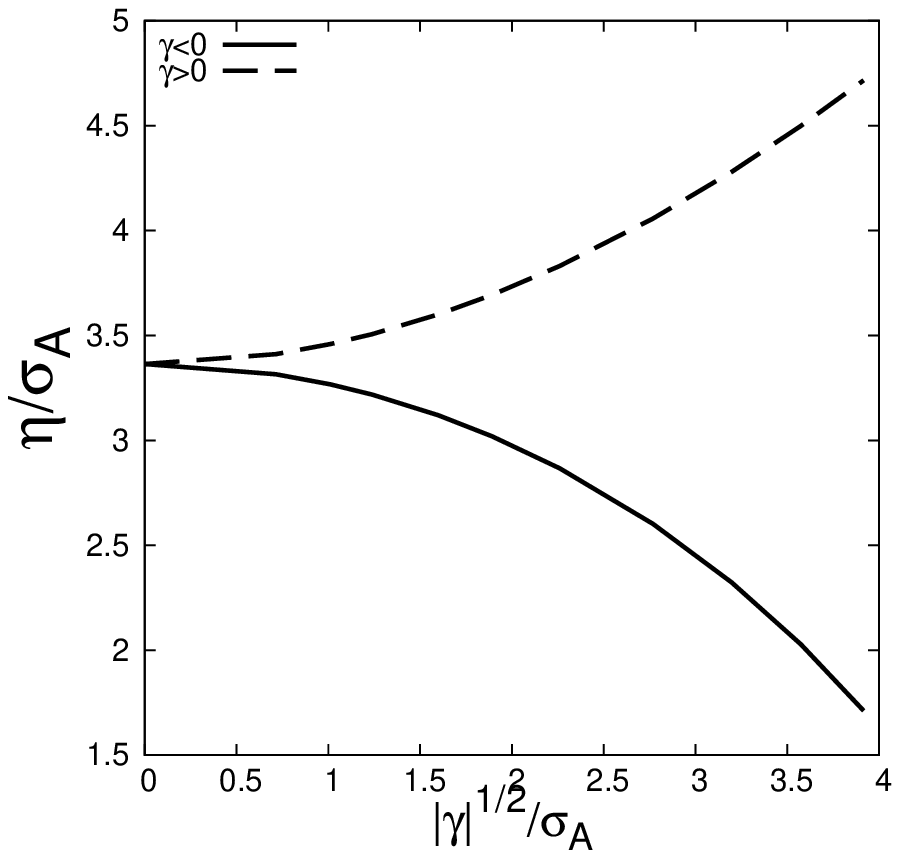}}
\end{center}
\caption{Eigenfrequency $\sigma/\sigma_A$ (left) and the growth rate $\eta/\sigma_A$ (right) of the $m=2$ magnetic modes versus the
  $|\gamma|^{1/2}/\sigma_A$ for the $n=1$ polytrope.}
\end{figure}

Using a dispersion relation derived by Lee (2010) for the oscillation of magnetized stars,
we try to explain the rapid spatial oscillations of the expansion coefficients $\pmb{H}$ and $\pmb{T}$ in the surface layers
for stable magnetic modes.
For a non-rotating and isentropic star, the dispersion relation may be given by
\begin{eqnarray}
-pq^2\cos^4\alpha\left(Rk\right)^6+q^2\cos^2\alpha\left(2{p\over q}+1\right)\bar\sigma^2\left(Rk\right)^4
+q\left[\bar\beta^2\cos^2\alpha-\left({p\over q}+1+\cos^2\alpha\right)\sigma^2\right]\bar\sigma^2\left(Rk\right)^2\nonumber\\
+\bar\sigma^2\left[\bar\sigma^4-\bar\sigma^2\bar\beta^2+q\left(Rk_H\right)^2\bar\beta^2\sin^2\psi\right]=0,
\label{eq:disp}
\end{eqnarray}
where
\be
p={a^2\over (R\Omega_K)^2}, \quad q={B^2/4\pi\rho\over (R\Omega_K)^2}, \quad \bar\beta^2={g^2/a^2\over \Omega_K^2},
\quad a^2=\Gamma_1{p\over\rho}, \quad g={GM_r\over r^2}, \quad \cos\alpha={\pmb{k}\cdot\pmb{B}\over kB}, \quad
\sin\psi={(\pmb{k}\times\pmb{B})_z\over k_HB_H},
\ee
and $k=|\pmb{k}|$ and $B=|\pmb{B}|$.
When we employ local cartesian coordinates $(x,y,z)$ and assume that
the $z$-, $x$-, and $y$-directions at a point in the interior are respectively along the $r$-, $\theta$-, and $\phi$-directions at that point,
we have $B_z=2f\cos\theta$, $B_y=0$, $B_x=-(2f+df/d\ln r)\sin\theta$, $k_H=\sqrt{k_x^2+k_y^2}$, $B_H=|B_x|$, and $\sin\psi=-k_yB_x/k_H|B_x|$.
Assuming a polytrope of index $n=1$ with the mass $M=1.4M_\odot$ and the radius $R=10^6$cm and
the field strength $B_S=10^{15}$G, we can compute the
quantities $p$, $q$, $\bar\beta^2$, $a^2$, $g$ as a function of $r$.
If we further assume for simplicity $k_xR=k_yR=1$ for $|m|\sim l\sim 1$
in the surface regions, we may solve equation (\ref{eq:disp}) for $(Rk)^2$ to obtain $(Rk_z)^2$ for given values of $\bar\sigma^2$, $\cos\theta$, and $\cos\alpha$.
Figure 10 shows the local wavelengths $\lambda/R=2\pi/(Rk_z)$ computed using the largest positive
solution $(Rk_z)^2$ in the outer envelope for several sets of parameters $(\bar\sigma^2,\cos\theta,\cos\alpha)$.
As shown by the figure, the waves
propagating across the field lines, which may be provided by the closed field lines near the surface, tend to have short wavelengths, which become even shorter toward the surface.
The wavelengths of the rapid spatial oscillations of the functions $\pmb{H}$ and $\pmb{T}$ near the surface
become comparable to those expected for the case of $\cos\alpha\sim 0.01$.
For low frequency modes we may expect $\pmb{k}\cdot\pmb{\xi}\sim0$, and hence 
$|\xi_z|/|\pmb{\xi}_H|\ll 1$ for $|k_z|/k_H\gg 1$.
In this case,  
rapid spatial oscillations in the radial components $\pmb{S}$ in the surface layers may not be conspicuous.

\begin{figure}
\begin{center}
\resizebox{0.4\columnwidth}{!}{
\includegraphics{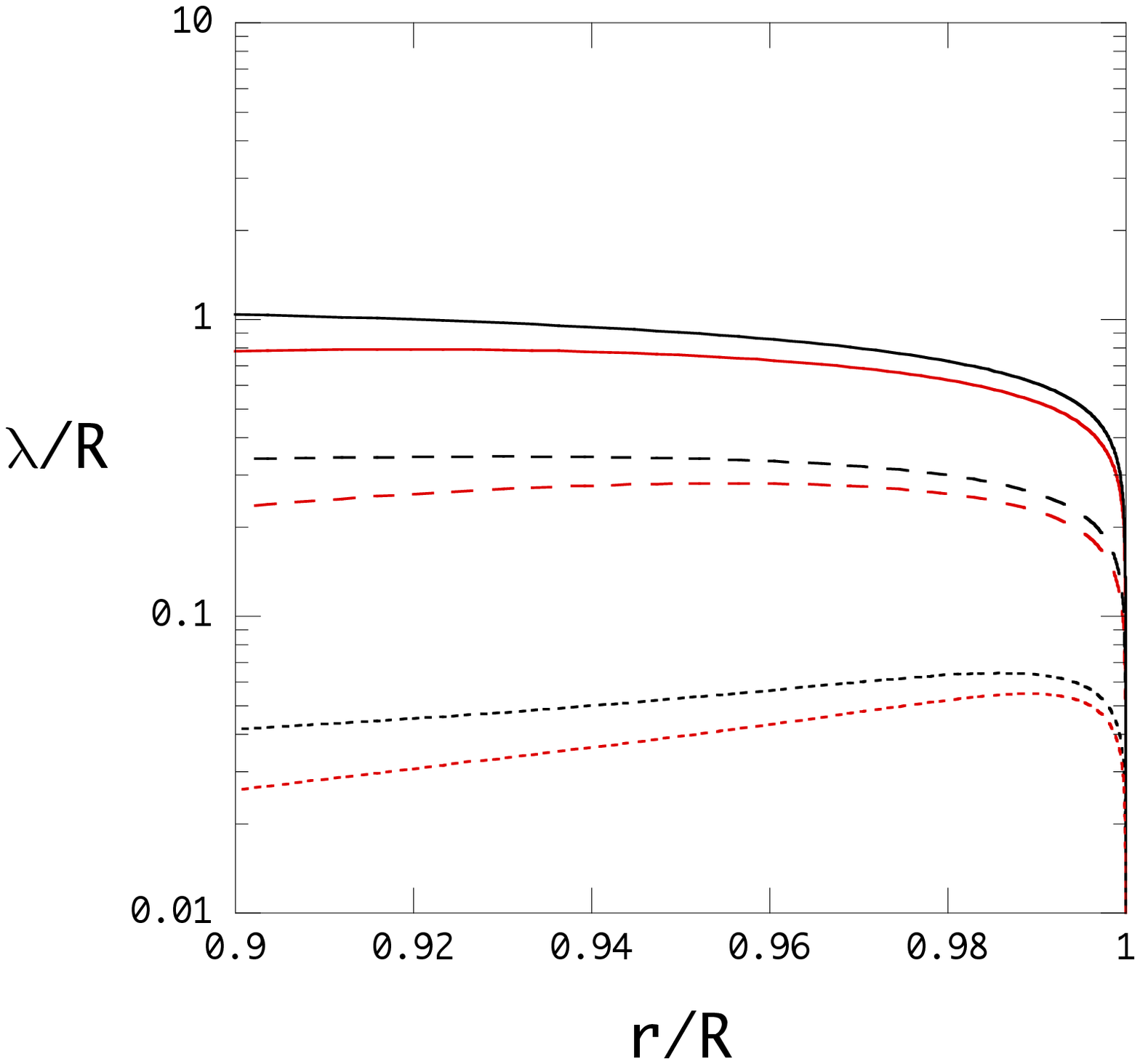}}
\hspace*{0.75cm}
\resizebox{0.4\columnwidth}{!}{
\includegraphics{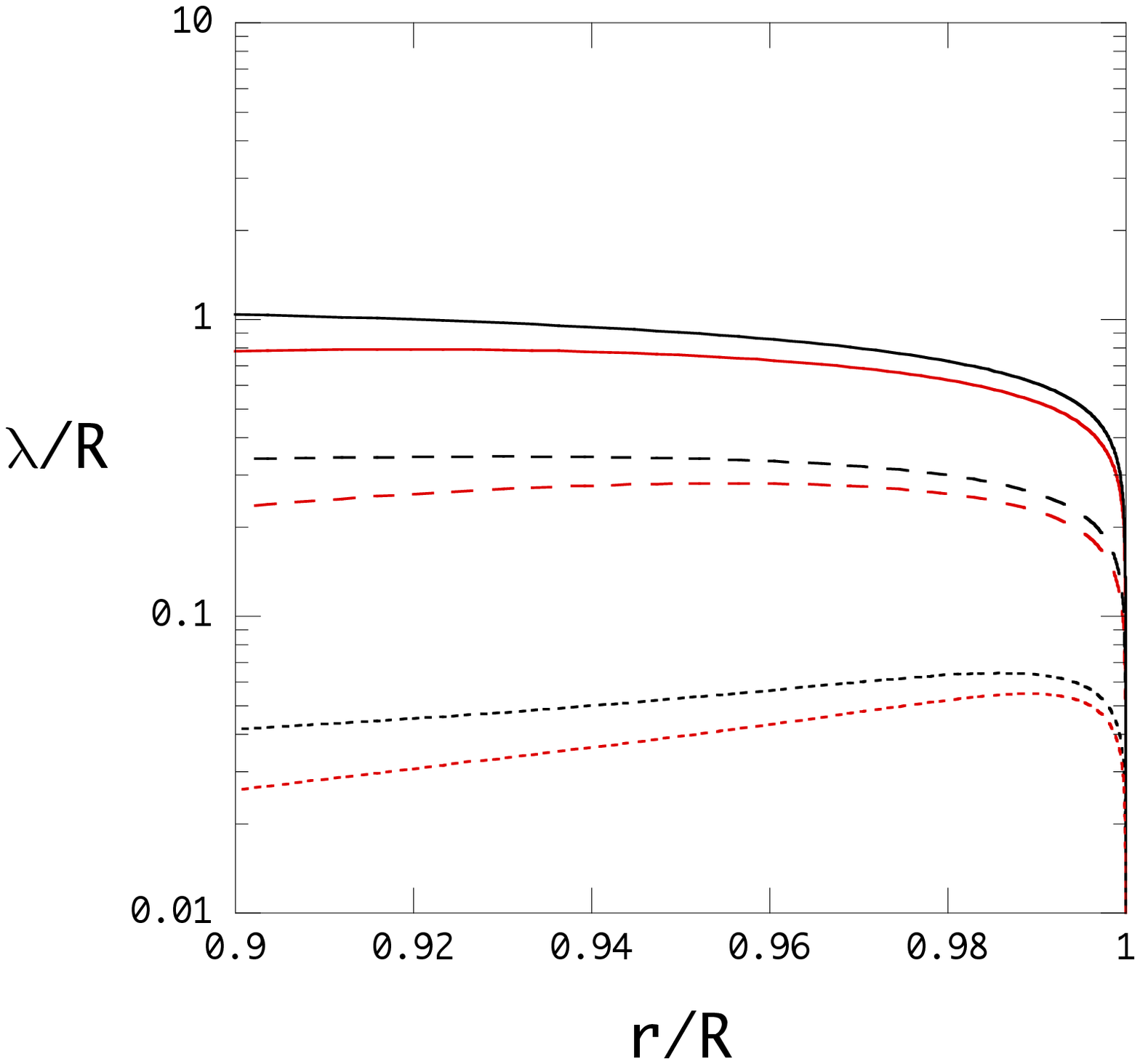}}
\end{center}
\caption{Local wavelengths $\lambda/R=2\pi/(Rk_z)$ derived using the dispersion relation (\ref{eq:disp}) for the $n=1$ polytrpoe of $M=1.4M_\odot$ and $R=10^6$cm and for $B_S=10^{15}$G, where the black lines and red lines are for $\cos\theta=0.5$ and
0, respectively, and the solid lines, dashed lines, and dotted lines are for $\cos\alpha=0.5$, 0.1, and 0.01,
respectively. Here, we assume $\bar\sigma^2=10^{-5}$ (left) and $\bar\sigma^2=10^{-4}$ (right).
}
\end{figure}

\section{Conclusion}

In this paper, 
we have calculated non-axisymmetric ($m\not=0$) oscillations of
neutron stars magnetized with purely poloidal magnetic fields, where we have
used polytropes of index $n=1$ and 1.5 as background neutron star models. 
We have found stable (oscillatory) magnetic modes ($\sigma^2>0$)
of odd parity and unstable (monotonically growing) magnetic modes ($\sigma^2<0$) of both even and odd parity.
The frequency $\sigma$ of the stable magnetic modes and the growth rate $\eta\equiv \sqrt{-\sigma^2}$ 
of the unstable magnetic modes are proportional to the magnetic field
strength $B_{\rm S}=\mu_b/R^3$ measured at the surface, if the effects of buoyancy
in the interior is negligible.
For a given $m$, the frequency $\sigma$ and the growth rate $\eta$ decrease as the number of radial nodes of the
eigenfunctions increases, which may indicate an anti-Strumian property of the problem.
We have found that the non-axisymmetric magnetic modes are affected by
stratification in the interior of the star, which is parametrized by using a parameter $\gamma$
in this paper.
For radiative stars ($\gamma<0$), the eigenfrequency $\sigma$ of the oscillatory magnetic modes
gradually increases as $|\gamma|$ increases, while it 
decreases with $|\gamma|$ for convective stars ($\gamma>0$).
For the unstable magnetic modes, on the other hand, we
have found that stable stratification with $\gamma<0$ reduces the growth rate $\eta$ of
the magnetic modes, while the convectively unstable stratification enhances the
growth rate as $|\gamma|$ is increased.
It is also found that the growth rate $\eta$ tends to increase as the
azimuthal number increases. We note that Lasky et al. (2011) obtained
strong instability of magnetized star having purely poloidal magnetic field, especially for $m=4$. 
It has been analytically shown that purely poloidal magnetic fields having closed field lines inside the star are 
unstable in the limit of $m\rightarrow\infty$ (see, e.g., van Assche, Goossens, \& Tayler 1982,  Markey \& Tayler 1973). 

As mentioned in \S4, we fail to obtain $f$- and $p$-modes of the magnetized star by using the present 
numerical code. At first glance, this fact might imply that the present numerical code and/or formulation 
have some problem. However, the situation at hand is not very simple, as we will discuss below. Let us 
assume that in the limit of $|\pmb{B}| \rightarrow 0$, there exist eigensolutions for  magnetized stars given by
\begin{eqnarray}
\sigma=\sigma_0+O(\bar{\sigma}_A^2) \,,\quad
\pmb{\xi}=\pmb{\xi}_0+O(\bar{\sigma}_A^2) \,,\quad p'=p'_0+O(\bar{\sigma}_A^2) \,,\quad 
\rho'=\rho'_0+O(\bar{\sigma}_A^2) \,,
\end{eqnarray}
where $\sigma_0$ is an eigenfrequency and $\pmb{\xi}_0$, $p'_0$, and $\rho'_0$ are its eigenfunctions for non-magnetized stars.  Clearly, Eq. (23) is 
an eigensolution of Eqs. (5)--(7)  in conjunction with the surface boundary condition $\Delta p(r=R)=0$ in  the limit of 
$|\pmb{B}| \rightarrow 0$. From Eq. (8), we obtain 
\begin{eqnarray}
{\pmb{B}'\over B_S}=\nabla\times\left(\pmb{\xi}_0\times{\pmb{B}\over B_S}\right)+O(\bar{\sigma}_A^2) \,. 
\end{eqnarray}
However, in general, this $\displaystyle {\pmb{B}^\prime\over B_S}$ does not satisfy the surface boundary condition that guarantees 
$\displaystyle {\pmb{B}^\prime\over B_S}\rightarrow 0$ as $r\rightarrow \infty$. Thus, Eqs. (23) and (24) cannot 
be an eigensolution for magnetized stars in the limit of $|\pmb{B}| \rightarrow 0$. In other words, oscillation modes that exist in 
non-magnetized stars need not be present in weakly magnetized stars. From physical point of view, however, it is plausible 
to expect that $f$- and $p$-modes exist in magnetized stars. 
This apparent contradiction may be resolved if we admit that in the normal mode analysis such as we carry out in this paper
the Lorentz force term in Eq. (5) must not 
vanish even in the limit of $|\pmb{B}| \rightarrow 0$.
This occurs if $\nabla\times\pmb{B}^\prime \propto \bar{\sigma}_A^{-1}$ 
is satisfied somewhere inside the star when $|\pmb{B}| \rightarrow 0$. 
That is to say, $\pmb{B}^\prime$ needs to show rapid spatial 
oscillations somewhere inside the star to keep the Lorentz terms finite as $|\pmb{B}| \rightarrow 0$. 
If this is the case, $f$- and $p$-modes of magnetized stars have to have slightly 
different eigenfrequencies and eigenfunctions form those of non-magnetized stars even in the limit of $|\pmb{B}| \rightarrow 0$. 
In the present numerical code, it will be difficult to treat the terms related to $\nabla\times\pmb{B}^\prime \propto \bar{\sigma}_A^{-1}$ 
accurately because we use a standard second-order accuracy finite-difference method for the radial direction and a spectral method for 
the $\theta$-direction, and some 
special technique will be required to evaluate the terms related to $\nabla\times\pmb{B}^\prime \propto \bar{\sigma}_A^{-1}$ properly. 
Actually, we find that we can obtain \lq$f$\rq- and \lq$p$\rq-modes if we largely reduce the number of mesh points distributed in
the interior for integration.
This reduction in the number of mesh points may be equivalent to averaging the rapid spatial oscillations of $\pmb{B}'$ over a length scale much larger than the wavelengths of the spatial oscillations, which avoids numerical difficulty associated with rapid spatial oscillations of $\pmb{B}'$.
However, we cannot consider that \lq$f$\rq- and \lq$p$\rq-modes thus obtained are correctly computed normal modes.

It is also important to note that, for given values of $m$ and $B_S$, we find only one magnetic mode sequence along the number of radial nodes of the eigenfunction.
This result is different from that for the axisymmetric toroidal magnetic modes of stars with a poloidal field as discussed by
Lee (2008) and Asai \& Lee (2014), who obtained several mode sequences, differing in the surface oscillation patterns, for given $m$ and $B_S$.
If discrete magnetic modes can exist only in the gaps between continuum frequency spectra, the difference in
the distribution of
continuum frequency spectra between axisymmetric $(m=0)$ toroidal modes and non-axisymmetric $(m\not=0)$ spheroidal modes, for given $m$ and $B_S$, might lead to the difference in the distribution of 
discrete magnetic modes between the two cases.

The present analysis is a part of our study of the oscillation of
magnetized stars. 
It is well known that a purely poloidal magnetic
field configuration is unstable as exemplified in this paper. 
Therefore, it may be difficult to observe long-lived
magnetic oscillations as QPOs. However a mixed poloidal and toroidal
magnetic field configuration such as twisted-torus magnetic field
(e.g., Braithwaite \& Spruit 2004; Yoshida \& Eriguchi 2006; Yoshida,
Yoshida \& Eriguchi 2006; Ciolfi et al. 2009) could be stable. 
Thus, it is interesting to investigate stability and
oscillation spectrum of the star having such a mixed magnetic field
configuration. 
In the presence of
both poloidal and toroidal field components, toroidal and spheroidal
modes are coupled even for axisymmetric perturbations, which inevitably
makes the analysis difficult.

As important physical properties inherent to cold neutron
stars, we need to consider the effects of solid crust and 
those of superfluidity and super-conductivity of neutrons and protons on the oscillation modes. 
It is believed that
neutrons become a superfluid both in the inner crust and in the fluid
core while protons can be superconducting in the core. For example, if
the fluid core is a type I superconductor, magnetic fields will be
expelled from the core region bacause of the Meissner effect, and
hence confined to the solid crust (e.g., Colaiuda et al. 2008; Sotani
et al. 2008). However, a recent analysis of the spectrum of
timing noise for SGR 1806-20 and SGR 1900+14 has suggested that the
core region is a type II superconductor (Arras, Cumming \& Thompson
2004). If this is the case, the fluid core can be threaded by a
magnetic field and hence the frequency spectra of oscillation modes
will be affected by the superconductivity in the core (e.g., Colaiuda
et al. 2008; Sotani et al. 2008). 
To investigate the oscillation of magnetized stars as normal modes, taking account of the effect of
superfluidity or superconductivity, will be one of our future studies
(see, e.g., Glampedakis, Andersson \& Samuelsson
2011; Passamonti \& Lander 2013, 2014; and Gabler et al. 2013b).

\appendix
\section{Pulsation equations for the magnetized star with purely poloidal magnetic fields}
To describe the master equations concisely, it is useful to introduce
the following column vectors composed of the expansion coefficients
for the perturbation quantities: the vectors $\pmb{S}$, $\pmb{H}$,
$\pmb{T}$, $\pmb{b}^S$, $\pmb{b}^H$, $\pmb{b}^T$, and $\delta\pmb{U}$,
defined by
\begin{eqnarray}
(\pmb{S})_j=S_{l_j}, \quad (\pmb{H})_j=H_{l_j}, \quad
(\pmb{T})_j=T_{l_j^\prime}, \quad (\pmb{b}^S)_j=b^S_{l_j^\prime},
\quad (\pmb{b}^H)_j=b^H_{l_j^\prime}, \quad (\pmb{b}^T)_j=b^T_{l_j},
\quad (\delta\pmb{U})_j=\frac{p^\prime_{l_j}}{\rho gr},
\end{eqnarray} 
where $(\pmb{X})_j$ denotes the $j$-th component of the vector
$\pmb{X}$ and $g=GM(r)/r^2$ is the gravitational acceleration. Here, 
$l_j=|m|+2(j-1)$ and $l_j^\prime=l_j+1$ for even modes, and
$l_j=|m|+2j-1$ and $l_j^\prime=l_j-1$ for odd modes, respectively, and
$j=1,2,3,...,j_{\rm max}$. 
The perturbed continuity equation (5) and the perturbed Euler equation (4) are then reduced to
\begin{eqnarray}
r\dfn{\pmb{S}}{r}{}=\left(\frac{V}{\Gamma_1}-3\right)\pmb{S}-\frac{V}{\Gamma_1}\delta\pmb{U}+\pmbmt{\Lambda}_0\pmb{H},
\end{eqnarray}
\begin{eqnarray}
-\frac{4\pi p}{rf}Vr\dfn{\delta\pmb{U}}{r}{}+\frac{4\pi
  p}{rf}V(c_1\bar\sigma^2+rA)\pmb{S}+\frac{4\pi
  p}{rf}V(1-rA-U)\delta\pmb{U} \nonumber \\
+\pmbmt{C}_0\left[(\rd f+2)r\dfn{\pmb{b}^H}{r}{}-(\rd
  f+2)\pmb{b}^S+(\rd^2f+6\rd f+4)\pmb{b}^H\right]+m\left[(\rd
  f+2)r\dfn{\pmb{b}^T}{r}{}+(\rd^2f+6\rd f+4)\pmb{b}^T\right]=0,
\end{eqnarray}
\begin{eqnarray}
\frac{V}{\Gamma_1}\delta\pmb{U}=\frac{V}{\Gamma_1}c_1\bar\sigma^2(\rd
f+2)\pmbmt{A}_1^{-1}\pmbmt{B}_1\pmb{S}-\frac{V}{\Gamma_1}mc_1\bar\sigma^2(\pmbmt{A}_1^{-1}+\pmbmt{\Lambda}_0^{-1}\pmbmt{A}_1^{-1}\pmbmt{\Lambda}_1)\pmb{T}+\bigg\{\frac{V}{\Gamma_1}c_1\bar\sigma^2\frac{r}{2f}\pmbmt{A}_1^{-1}
\nonumber \\
+\frac{rf}{2\pi\Gamma_1p}\left[(\rd^2f+4\rd
  f)\pmbmt{\Lambda}_0^{-1}\pmbmt{B}_0-\pmbmt{\Lambda}_0^{-1}\pmbmt{A}_0+\frac{1}{2}m^2(\rd^2f+4\rd
  f+2)\pmbmt{\Lambda}_0^{-1}\pmbmt{A}_1^{-1}\right]\bigg\}\pmb{b}^S
\nonumber \\
+\frac{rf}{\pi\Gamma_1p}(\pmbmt{\Lambda}_0^{-1}\pmbmt{A}_0-m^2\pmbmt{\Lambda}_0^{-1}\pmbmt{A}_1^{-1})\pmb{b}^H+\frac{rf}{2\pi\Gamma_1p}m\bigg[\frac{1}{2}(\rd
f+2)\pmbmt{I}+2\pmbmt{\Lambda}_0^{-1} \nonumber \\
\rd f\pmbmt{\Lambda}_0^{-1}\pmbmt{A}_1^{-1}\pmbmt{B}_1\pmbmt{\Lambda}_0+\pmbmt{\Lambda}_0^{-1}\pmbmt{A}_1^{-1}\tilde{\pmbmt{C}}_1\bigg]\pmb{b}^T+\frac{rf}{2\pi\Gamma_1p}\pmbmt{\Lambda}_0^{-1}\pmbmt{A}_0\pmbmt{L}_0r\dfn{\pmb{b}^H}{r}{},
\end{eqnarray}
\begin{eqnarray}
\pmbmt{A}_1r\dfn{\pmb{b}^T}{r}{}=-\frac{2\pi
  p}{rf}Vc_1\bar\sigma^2\pmbmt{\Lambda}_1\pmb{T}+\frac{1}{2}m(\rd^2f+4\rd
f+2)\pmb{b}^S-2m\pmb{b}^H+\left(\rd f\pmbmt{B}_1\pmbmt{\Lambda}_0+\tilde{\pmbmt{C}}_1\right)\pmb{b}^T-mr\dfn{\pmb{b}^H}{r}{}. 
\end{eqnarray}
Substituting $\delta\pmb{U}$ given in (A4) into (A3), we may obtain 
\begin{eqnarray}
\pmbmt{L}_0r\dfn{}{r}{}\left(r\dfn{\pmb{b}^H}{r}{}\right)=\frac{2\pi
  p}{rf}V\left(c_1\bar{\sigma}^2+rA\right)\pmbmt{A}_0^{-1}\pmbmt{\Lambda}_0\pmb{S}+\frac{4\pi
  p}{rf}mVc_1\bar{\sigma}^2\pmbmt{A}_0^{-1}\pmbmt{A}_1^{-1}\pmbmt{\Lambda}_1\pmb{T}+\bigg\{-\frac{1}{2}\left(\rd
  f+2\right)\pmbmt{A}_0^{-1}\tilde{\pmbmt{C}}_0-\left(\rd\rho-rA+4\right)\pmbmt{I}
\nonumber \\
+\left[-\rd^3f+\rd^2f\left(\rd\rho-rA-4\right)+4\rd
  f\left(\rd\rho-rA+1\right)\right]\pmbmt{A}_0^{-1}\pmbmt{B}_0-\frac{1}{2}\big[\rd^3f-\rd^2f(\rd\rho-4)
\nonumber \\
-2\rd
f(2\rd\rho+1)-2(\rd\rho+2)\big]m^2\pmbmt{A}_0^{-1}\pmbmt{A}_1^{-1}\bigg\}\pmb{b}^S+\bigg[(\rd
f+1)\pmbmt{A}_0^{-1}\tilde{\pmbmt{C}}_0+\pmbmt{\Lambda}_1+2\pmbmt{A}_0^{-1}\pmbmt{B}_0\pmbmt{\Lambda}_1+2(\rd\rho-rA)\pmbmt{I}
\nonumber \\
-\frac{1}{2}(\rd^2f+4\rd
f+2)m^2\pmbmt{A}_0^{-1}\pmbmt{A}_1^{-1}\tilde{\pmbmt{\Lambda}}_1-(\rd^2f+2\rd
f+2\rd\rho)m^2\pmbmt{A}_0^{-1}\pmbmt{A}_1^{-1}\bigg]\pmb{b}^H
\nonumber \\
+\Bigg\{-m\bigg\{2(\rd f-\rd\rho+rA+1)\pmbmt{A}_0^{-1}+\frac{1}{2}[-\rd
f(\rd\rho+2)+(\rd
f+2)rA-2(\rd\rho+1)]\pmbmt{A}_0^{-1}\pmbmt{\Lambda}_0\bigg\} \nonumber
\\
-m\left[(\rd^2f-\rd\rho\rd f+2\rd
  f)\pmbmt{A}_0^{-1}\pmbmt{A}_1^{-1}\pmbmt{B}_1\pmbmt{\Lambda}_0+(\rd
  f-\rd\rho+1)\pmbmt{A}_0^{-1}\pmbmt{A}_1^{-1}\tilde{\pmbmt{C}}_1\right]\Bigg\}\pmb{b}^T
\nonumber \\
+\left[\frac{1}{2}(\rd
  f+2)\pmbmt{A}_0^{-1}\tilde{\pmbmt{C}}_0+(\rd\rho-rA-1)\pmbmt{I}+m^2(\rd
  f-\rd\rho+3)\pmbmt{A}_0^{-1}\pmbmt{A}_1^{-1}\right]r\dfn{\pmb{b}^H}{r}{}-\frac{2\pi
  p}{rf}(2+rA)Vc_1\bar{\sigma}^2\pmbmt{A}_0^{-1}\pmbmt{\Lambda}_0\pmb{H}
\nonumber \\
-\frac{2\pi
  p}{rf}Vc_1\bar{\sigma}^2\pmbmt{A}_0^{-1}\pmbmt{\Lambda}_0r\dfn{\pmb{H}}{r}{}+\frac{2\pi
  p}{rf}mVc_1\bar{\sigma}^2\pmbmt{A}_0^{-1}\pmbmt{A}_1^{-1}\pmbmt{\Lambda}_1r\dfn{\pmb{T}}{r}{}
-m\left[\frac{1}{2}(\rd
  f+2)\pmbmt{A}_0^{-1}\pmbmt{A}_1^{-1}\tilde{\pmbmt{C}}_1+(\rd f+rA+2)\pmbmt{A}_0^{-1}\right]r\dfn{\pmb{b}^T}{r}{}.
\end{eqnarray}
The perturbed induction equation (7) and the perturbed Gauss's law for magnetic fields are reduced to
\begin{eqnarray}
\pmb{H}=(\rd f+2)\pmbmt{A}_1^{-1}\pmbmt{B}_1\pmb{S}-m\pmbmt{A}_1^{-1}\pmb{T}+\frac{r}{2f}\pmbmt{A}_1^{-1}\pmb{b}^S,
\end{eqnarray}
\begin{eqnarray}
\pmbmt{L}_0r\dfn{\pmb{T}}{r}{}=-m\left[\rd^2f+\left(2+\frac{V}{\Gamma_1}\right)\rd
  f-2\left(1-\frac{V}{\Gamma_1}\right)\right]\left(\pmbmt{A}_0^{-1}\pmbmt{A}_1^{-1}\pmbmt{B}_1+\frac{1}{2}\pmbmt{A}_0^{-1}\right)\pmb{S}+\left(\rd
  f+2\right)\left(m^2\pmbmt{A}_0^{-1}\pmbmt{A}_1^{-1}+\frac{1}{2}\pmbmt{A}_0^{-1}\tilde{\pmbmt{C}}_0\right)\pmb{T}
\nonumber \\
-\frac{r}{2f}m\pmbmt{A}_0^{-1}\pmbmt{A}_1^{-1}\pmbmt{\Lambda}_1\pmb{b}^H+\frac{r}{2f}\pmbmt{A}_0^{-1}\pmbmt{\Lambda}_0\pmb{b}^T-m\left(\rd
  f+2\right)\left(\pmbmt{A}_0^{-1}+\frac{1}{2}\pmbmt{A}_0^{-1}\pmbmt{A}_1^{-1}\tilde{\pmbmt{C}}_1\right)\pmb{H}
\nonumber \\
+m\left(\rd f+2\right)\left(\pmbmt{A}_0^{-1}\pmbmt{A}_1^{-1}\pmbmt{B}_1+\frac{1}{2}\pmbmt{A}_0^{-1}\right)\frac{V}{\Gamma_1}\delta\pmb{U},
\end{eqnarray}
\begin{eqnarray}
\pmbmt{L}_1r\dfn{\pmb{H}}{r}{}=\left[\rd^2f+\left(2+\frac{V}{\Gamma_1}\right)\rd
  f-2\left(1-\frac{V}{\Gamma_1}\right)\right]\left(\pmbmt{A}_1^{-1}\pmbmt{B}_1+\frac{1}{2}m^2\pmbmt{A}_1^{-1}\pmbmt{A}_0^{-1}\right)\pmb{S}-m\left(\rd
  f+2\right)\left(\pmbmt{A}_1^{-1}+\frac{1}{2}\pmbmt{A}_1^{-1}\pmbmt{A}_0^{-1}\tilde{\pmbmt{C}}_0\right)\pmb{T}
\nonumber \\
+\frac{r}{2f}\pmbmt{A}_1^{-1}\pmbmt{\Lambda}_1\pmb{b}^H-\frac{r}{2f}m\pmbmt{A}_1^{-1}\pmbmt{A}_0^{-1}\pmbmt{\Lambda}_0\pmb{b}^T+\left(\rd
  f+2\right)\left(m^2\pmbmt{A}_1^{-1}\pmbmt{A}_0^{-1}+\frac{1}{2}\pmbmt{A}_1^{-1}\tilde{\pmbmt{C}}_1\right)\pmb{H}
\nonumber \\
-\left(\rd f+2\right)\left(\pmbmt{A}_1^{-1}\pmbmt{B}_1+\frac{1}{2}m^2\pmbmt{A}_1^{-1}\pmbmt{A}_0^{-1}\right)\frac{V}{\Gamma_1}\delta\pmb{U},
\end{eqnarray}
\begin{eqnarray}
r\dfn{\pmb{b}^S}{r}{}=-3\pmb{b}^S+\pmbmt{\Lambda}_1\pmb{b}^H.
\end{eqnarray}
Here
\begin{eqnarray}
V=-\dfn{\ln p}{\ln r}{}, \quad
c_1=\frac{M}{M(r)}\left(\frac{r}{R}\right)^3, \quad
\rd\rho=\dfn{\ln\rho}{\ln r}{}, \quad \rd f=\dfn{\ln f}{\ln r}{},
\quad \rd^2f=\frac{r^2}{f}\dfn{f}{r}{2}, \quad \rd^3f=\frac{r^3}{f}\dfn{f}{r}{3},
\end{eqnarray}
and $\bar\sigma\equiv\sigma/(GM/R^3)^{1/2}$ is the frequency in the
unit of the Kepler frequency of the star. The quantities $\pmbmt{A}_0$,
$\pmbmt{A}_1$, $\pmbmt{B}_0$, $\pmbmt{B}_1$, $\tilde{\pmbmt{C}}_0$,
$\tilde{\pmbmt{C}}_1$, $\tilde{\pmbmt{\Lambda}}_0$,
$\tilde{\pmbmt{\Lambda}}_1$, $\pmbmt{L}_0$, and $\pmbmt{L}_1$ denote
the matrices defined as follows:
\begin{eqnarray}
\pmbmt{A}_0=\pmbmt{C}_0+\pmbmt{Q}_0\pmbmt{\Lambda}_1,\quad
\pmbmt{A}_1=\pmbmt{C}_1+\pmbmt{Q}_1\pmbmt{\Lambda}_0,\quad
\pmbmt{B}_0=\pmbmt{Q}_0+\frac{1}{2}\pmbmt{C}_0,\quad
\pmbmt{B}_1=\pmbmt{Q}_1+\frac{1}{2}\pmbmt{C}_1, \nonumber
\end{eqnarray}
\begin{eqnarray}
\tilde{\pmbmt{C}}_0=\pmbmt{C}_0(\pmbmt{\Lambda}_1-2\pmbmt{I}),\quad
\tilde{\pmbmt{C}}_1=\pmbmt{C}_1(\pmbmt{\Lambda}_0-2\pmbmt{I}),\quad
\tilde{\pmbmt{\Lambda}}_0=\pmbmt{\Lambda}_0-2\pmbmt{I},\quad
\tilde{\pmbmt{\Lambda}}_1=\pmbmt{\Lambda}_1-2\pmbmt{I}, \nonumber
\end{eqnarray}
\begin{eqnarray}
\pmbmt{L}_0=\pmbmt{I}-m^2\pmbmt{A}_0^{-1}\pmbmt{A}_1^{-1},\quad \pmbmt{L}_1=\pmbmt{I}-m^2\pmbmt{A}_1^{-1}\pmbmt{A}_0^{-1}.
\end{eqnarray}
The matrices $\pmbmt{Q}_0$, $\pmbmt{Q}_1$, $\pmbmt{C}_0$,
$\pmbmt{C}_1$, $\pmbmt{\Lambda}_0$, and $\pmbmt{\Lambda}_1$ are
defined as follows:

\noindent
For even modes, 
\begin{eqnarray}
(\pmbmt{Q}_0)_{jj}=J_{l_j+1}^m, \quad
(\pmbmt{Q}_0)_{j+1,j}=J_{l_j+2}^m,\quad
(\pmbmt{Q}_1)_{jj}=J_{l_j+1}^m,\quad
(\pmbmt{Q}_1)_{j,j+1}=J_{l_j+2}^m, \nonumber
\end{eqnarray}
\begin{eqnarray}
(\pmbmt{C}_0)_{jj}=-(l_j+2)J_{l_j+1}^m,\quad
(\pmbmt{C}_0)_{j+1,j}=(l_j+1)J_{l_j+2}^m, \quad
(\pmbmt{C}_1)_{jj}=l_jJ_{l_j+1}^m,\quad
(\pmbmt{C}_1)_{j,j+1}=-(l_j+3)J_{l_j+2}^m, \nonumber
\end{eqnarray}
\begin{eqnarray}
(\pmbmt{\Lambda}_0)_{jj}=l_j(l_j+1),\quad
(\pmbmt{\Lambda}_1)_{jj}=(l_j+1)(l_j+2),
\end{eqnarray}
where $l_j=|m|+2j-2$ for $j=1,2,3,....,j_{\rm max}$, and 
\begin{eqnarray}
J_{l_j}^m=\left[\frac{(l_j+m)(l_j-m)}{(2l_j-1)(2l_j+1)}\right]^{1/2}.
\end{eqnarray}
For odd modes,
\begin{eqnarray}
(\pmbmt{Q}_0)_{jj}=J_{l_j+1}^m,\quad
(\pmbmt{Q}_0)_{j,j+1}=J_{l_j+2}^m, \quad
(\pmbmt{Q}_1)_{jj}=J_{l_j+1}^m,\quad
(\pmbmt{Q}_1)_{j+1,j}=J_{l_j+2}^m, \nonumber
\end{eqnarray}
\begin{eqnarray}
(\pmbmt{C}_0)_{jj}=l_jJ_{l_j+1}^m,\quad
(\pmbmt{C}_0)_{j,j+1}=-(l_j+3)J_{l_j+2}^m, \quad
(\pmbmt{C}_1)_{jj}=-(l_j+2)J_{l_j+1}^m,\quad
(\pmbmt{C}_1)_{j+1,j}=(l_j+1)J_{l_j+2}^m, \nonumber
\end{eqnarray}
\begin{eqnarray}
(\pmbmt{\Lambda}_0)_{jj}=(l_j+1)(l_j+2),\quad
(\pmbmt{\Lambda}_1)_{jj}=l_j(l_j+1),
\end{eqnarray}
where $l_j=|m|+2j-1$ for $j=1,2,3,....,j_{\rm max}$.

From the equations given before, we see that 
non-axisymmetric pulsations of the magnetized star with purely poloidal magnetic fields may be described by a
system of $6j_{\rm max}$-th order ordinary differential equations. 
In this study, we chose  the column vectors $\pmb{S}$, $\pmb{T}$, $\pmb{b}^S$, $\pmb{b}^H$, $\pmb{b}^T$, and
$\displaystyle r{\rd\over \rd r}\pmb{b}^H$ as dependent variables. If the dimensionless vector variables, defined by
\begin{eqnarray}
\pmb{y}_1=\pmb{S}, \quad \pmb{y}_2=\pmb{T}, \quad \pmb{y}_3=\pmb{h}^S,
\quad \pmb{y}_4=\pmb{h}^H, \quad \pmb{y}_5=\pmb{h}^T, \quad \pmb{y}_6=r\dfn{}{r}{}\pmb{h}^H,
\end{eqnarray}
where $\pmb{h}^i\equiv [R/f(0)]\pmb{b}^i$ for $i=S, H, T$, are introduced, the master equations
for stellar pulsations are schematically written by the coupled first-order differential equations, given by 
\begin{eqnarray}
r\dfn{\pmb{y}_1}{r}{}=\pmbmt{{\cal{F}}}_{11}\pmb{y}_1+\pmbmt{\Lambda}_0\pmb{H}-\frac{V}{\Gamma_1}\delta\pmb{U} \ \ \ \ \ \ \ \ \ \ \ \ \ \ \ \ \ \ \ \ \ \ \ \ \ \ \ \ \ \ \ \ \ \ \ \ \ \ \ \ \ \ \ \ \ \ \ \ \ \ \ \ \ \ \ \ \ \ \ \ \ \ \ \ \ \ \ \ \ \ \ \ \ \ \ \ \ \ \ \ \ \ \ \ \ \ \ \ \ \nonumber \\
\ \ \ \ \ \ \ =(\pmbmt{{\cal{F}}}_{11}-\pmbmt{{\cal{E}}}_{11}+\pmbmt{\Lambda}_0\pmbmt{{\cal{E}}}_{21})\pmb{y}_1+(-\pmbmt{{\cal{E}}}_{12}+\pmbmt{\Lambda}_0\pmbmt{{\cal{E}}}_{22})\pmb{y}_2+(-\pmbmt{{\cal{E}}}_{13}+\pmbmt{\Lambda}_0\pmbmt{{\cal{E}}}_{23})\pmb{y}_3-\pmbmt{{\cal{E}}}_{14}\pmb{y}_4-\pmbmt{{\cal{E}}}_{15}\pmb{y}_5-\pmbmt{{\cal{E}}}_{16}\pmb{y}_6,
\end{eqnarray}
\begin{eqnarray}
r\dfn{\pmb{y}_2}{r}{}=\pmbmt{{\cal{F}}}_{21}\pmb{y}_1+\pmbmt{{\cal{F}}}_{22}\pmb{y}_2+\pmbmt{{\cal{F}}}_{24}\pmb{y}_4+\pmbmt{{\cal{F}}}_{25}\pmb{y}_5+\pmbmt{{\cal{G}}}_{21}\frac{V}{\Gamma_1}\delta\pmb{U}+\pmbmt{{\cal{G}}}_{22}\pmb{H} \ \ \ \ \ \ \ \ \ \ \ \ \ \ \ \ \ \ \ \ \ \ \ \ \ \ \ \ \ \ \ \ \ \ \nonumber \\
\ \ \ \ \ \  =(\pmbmt{{\cal{F}}}_{21}+\pmbmt{{\cal{G}}}_{21}\pmbmt{{\cal{E}}}_{11}+\pmbmt{{\cal{G}}}_{22}\pmbmt{{\cal{E}}}_{21})\pmb{y}_1+(\pmbmt{{\cal{F}}}_{22}+\pmbmt{{\cal{G}}}_{21}\pmbmt{{\cal{E}}}_{12}+\pmbmt{{\cal{G}}}_{22}\pmbmt{{\cal{E}}}_{22})\pmb{y}_2+(\pmbmt{{\cal{G}}}_{21}\pmbmt{{\cal{E}}}_{13}+\pmbmt{{\cal{G}}}_{22}\pmbmt{{\cal{E}}}_{23})\pmb{y}_3 \nonumber \\
+(\pmbmt{{\cal{F}}}_{24}+\pmbmt{{\cal{G}}}_{21}\pmbmt{{\cal{E}}}_{14})\pmb{y}_4+(\pmbmt{{\cal{F}}}_{25}+\pmbmt{{\cal{G}}}_{21}\pmbmt{{\cal{E}}}_{15})\pmb{y}_5+\pmbmt{{\cal{G}}}_{21}\pmbmt{{\cal{E}}}_{16}\pmb{y}_6,
\end{eqnarray}
\begin{eqnarray}
r\dfn{\pmb{y}_3}{r}{}=-3\pmb{y}_3+\pmbmt{\Lambda}_1\pmb{y}_4,
\end{eqnarray}
\begin{eqnarray}
r\dfn{\pmb{y}_4}{r}{}=\pmb{y}_6,
\end{eqnarray}
\begin{eqnarray}
r\dfn{\pmb{y}_5}{r}{}=\pmbmt{{\cal{F}}}_{52}\pmb{y}_2+\pmbmt{{\cal{F}}}_{53}\pmb{y}_3+\pmbmt{{\cal{F}}}_{54}\pmb{y}_4+\pmbmt{{\cal{F}}}_{55}\pmb{y}_5+\pmbmt{{\cal{F}}}_{56}\pmb{y}_6,
\end{eqnarray}
\begin{eqnarray}
r\dfn{\pmb{y}_6}{r}{}=\pmbmt{{\cal{F}}}_{61}\pmb{y}_1+\pmbmt{{\cal{F}}}_{62}\pmb{y}_2+\pmbmt{{\cal{F}}}_{63}\pmb{y}_3+\pmbmt{{\cal{F}}}_{64}\pmb{y}_4+\pmbmt{{\cal{F}}}_{65}\pmb{y}_5+\pmbmt{{\cal{F}}}_{66}\pmb{y}_6+\pmbmt{{\cal{G}}}_{62}\pmb{H}+\pmbmt{{\cal{G}}}_{63}r\dfn{\pmb{H}}{r}{}+\pmbmt{{\cal{G}}}_{64}r\dfn{\pmb{y}_2}{r}{}+\pmbmt{{\cal{G}}}_{65}r\dfn{\pmb{y}_5}{r}{} \nonumber \\
=\left[\pmbmt{{\cal{F}}}_{61}+\pmbmt{{\cal{G}}}_{62}\pmbmt{{\cal{E}}}_{21}+\pmbmt{{\cal{G}}}_{63}(\pmbmt{{\cal{E}}}_{31}+\pmbmt{{\cal{E}}}_{37}\pmbmt{{\cal{E}}}_{21}+\pmbmt{{\cal{E}}}_{38}\pmbmt{{\cal{E}}}_{11})+\pmbmt{{\cal{G}}}_{64}(\pmbmt{{\cal{F}}}_{21}+\pmbmt{{\cal{G}}}_{21}\pmbmt{{\cal{E}}}_{11}+\pmbmt{{\cal{G}}}_{22}\pmbmt{{\cal{E}}}_{21})\right]\pmb{y}_1 \nonumber \\
+\left[\pmbmt{{\cal{F}}}_{62}+\pmbmt{{\cal{G}}}_{62}\pmbmt{{\cal{E}}}_{22}+\pmbmt{{\cal{G}}}_{63}(\pmbmt{{\cal{E}}}_{32}+\pmbmt{{\cal{E}}}_{37}\pmbmt{{\cal{E}}}_{22}+\pmbmt{{\cal{E}}}_{38}\pmbmt{{\cal{E}}}_{12})+\pmbmt{{\cal{G}}}_{64}(\pmbmt{{\cal{F}}}_{22}+\pmbmt{{\cal{G}}}_{21}\pmbmt{{\cal{E}}}_{12}+\pmbmt{{\cal{G}}}_{22}\pmbmt{{\cal{E}}}_{22})+\pmbmt{{\cal{G}}}_{65}\pmbmt{{\cal{F}}}_{52}\right]\pmb{y}_2 \nonumber \\
+\left[\pmbmt{{\cal{F}}}_{63}+\pmbmt{{\cal{G}}}_{62}\pmbmt{{\cal{E}}}_{23}+\pmbmt{{\cal{G}}}_{63}(\pmbmt{{\cal{E}}}_{37}\pmbmt{{\cal{E}}}_{23}+\pmbmt{{\cal{E}}}_{38}\pmbmt{{\cal{E}}}_{13})+\pmbmt{{\cal{G}}}_{64}(\pmbmt{{\cal{G}}}_{21}\pmbmt{{\cal{E}}}_{13}+\pmbmt{{\cal{G}}}_{22}\pmbmt{{\cal{E}}}_{23})+\pmbmt{{\cal{G}}}_{65}\pmbmt{{\cal{F}}}_{53}\right]\pmb{y}_3 \nonumber \\
+\left[\pmbmt{{\cal{F}}}_{64}+\pmbmt{{\cal{G}}}_{63}(\pmbmt{{\cal{E}}}_{34}+\pmbmt{{\cal{E}}}_{38}\pmbmt{{\cal{E}}}_{14})+\pmbmt{{\cal{G}}}_{64}(\pmbmt{{\cal{F}}}_{24}+\pmbmt{{\cal{G}}}_{21}\pmbmt{{\cal{E}}}_{14})+\pmbmt{{\cal{G}}}_{65}\pmbmt{{\cal{F}}}_{54}\right]\pmb{y}_4 \nonumber \\
+\left[\pmbmt{{\cal{F}}}_{65}+\pmbmt{{\cal{G}}}_{63}(\pmbmt{{\cal{E}}}_{35}+\pmbmt{{\cal{E}}}_{38}\pmbmt{{\cal{E}}}_{15})+\pmbmt{{\cal{G}}}_{64}(\pmbmt{{\cal{F}}}_{25}+\pmbmt{{\cal{G}}}_{21}\pmbmt{{\cal{E}}}_{15})+\pmbmt{{\cal{G}}}_{65}\pmbmt{{\cal{F}}}_{55}\right]\pmb{y}_5 \nonumber \\
+\left(\pmbmt{{\cal{F}}}_{66}+\pmbmt{{\cal{G}}}_{63}\pmbmt{{\cal{E}}}_{38}\pmbmt{{\cal{E}}}_{16}+\pmbmt{{\cal{G}}}_{64}\pmbmt{{\cal{G}}}_{21}\pmbmt{{\cal{E}}}_{16}+\pmbmt{{\cal{G}}}_{65}\pmbmt{{\cal{F}}}_{56}\right)\pmb{y}_6, 
\end{eqnarray}
and the algebraic relations, given by 
\begin{eqnarray}
\frac{V}{\Gamma_1}\delta\pmb{U}=\pmbmt{{\cal{E}}}_{11}\pmb{y}_1+\pmbmt{{\cal{E}}}_{12}\pmb{y}_2+\pmbmt{{\cal{E}}}_{13}\pmb{y}_3+\pmbmt{{\cal{E}}}_{14}\pmb{y}_4+\pmbmt{{\cal{E}}}_{15}\pmb{y}_5+\pmbmt{{\cal{E}}}_{16}\pmb{y}_6,
\end{eqnarray}
\begin{eqnarray}
\pmb{H}=\pmbmt{{\cal{E}}}_{21}\pmb{y}_1+\pmbmt{{\cal{E}}}_{22}\pmb{y}_2+\pmbmt{{\cal{E}}}_{23}\pmb{y}_3,
\end{eqnarray}
\begin{eqnarray}
r\dfn{\pmb{H}}{r}{}=\pmbmt{{\cal{E}}}_{31}\pmb{y}_1+\pmbmt{{\cal{E}}}_{32}\pmb{y}_2+\pmbmt{{\cal{E}}}_{34}\pmb{y}_4+\pmbmt{{\cal{E}}}_{35}\pmb{y}_5+\pmbmt{{\cal{E}}}_{37}\pmb{H}+\pmbmt{{\cal{E}}}_{38}\frac{V}{\Gamma_1}\delta\pmb{U} \ \ \ \ \ \ \ \ \ \ \ \ \ \ \ \ \ \ \ \ \ \ \ \ \ \ \ \ \ \ \ \ \ \ \ \ \nonumber \\
\ \ \ \ =(\pmbmt{{\cal{E}}}_{31}+\pmbmt{{\cal{E}}}_{37}\pmbmt{{\cal{E}}}_{21}+\pmbmt{{\cal{E}}}_{38}\pmbmt{{\cal{E}}}_{11})\pmb{y}_1+(\pmbmt{{\cal{E}}}_{32}+\pmbmt{{\cal{E}}}_{37}\pmbmt{{\cal{E}}}_{22}+\pmbmt{{\cal{E}}}_{38}\pmbmt{{\cal{E}}}_{12})\pmb{y}_2+(\pmbmt{{\cal{E}}}_{37}\pmbmt{{\cal{E}}}_{23}+\pmbmt{{\cal{E}}}_{38}\pmbmt{{\cal{E}}}_{13})\pmb{y}_3 \nonumber \\
+(\pmbmt{{\cal{E}}}_{34}+\pmbmt{{\cal{E}}}_{38}\pmbmt{{\cal{E}}}_{14})\pmb{y}_4+(\pmbmt{{\cal{E}}}_{35}+\pmbmt{{\cal{E}}}_{38}\pmbmt{{\cal{E}}}_{15})\pmb{y}_5+\pmbmt{{\cal{E}}}_{38}\pmbmt{{\cal{E}}}_{16}\pmb{y}_6. 
\end{eqnarray}
The coefficient matrices appearing in (A17)-(A25) are defined by
\begin{eqnarray}
\pmbmt{{\cal{E}}}_{11}=V_Gc_1\bar{\sigma}^2\left(\rd f+2\right)\pmbmt{A}_1^{-1}\pmbmt{B}_1,\quad
\pmbmt{{\cal{E}}}_{12}=-mV_Gc_1\bar{\sigma}^2\left(\pmbmt{A}_1^{-1}+\pmbmt{\Lambda}_0^{-1}\pmbmt{A}_1^{-1}\pmbmt{\Lambda}_1\right), \nonumber
\end{eqnarray}
\begin{eqnarray}
\pmbmt{{\cal{E}}}_{13}=\frac{x}{2\hat{f}}V_Gc_1\bar{\sigma}^2\pmbmt{A}_1^{-1}+\frac{\hat{f}}{2\hat{\rho}x}V_Gc_1\bar{\omega}^2_A\left[\left(\rd^2f+4\rd f\right)\pmbmt{\Lambda}_0^{-1}\pmbmt{B}_0-\pmbmt{\Lambda}_0^{-1}\pmbmt{A}_0+\frac{1}{2}m^2\left(\rd^2f+4\rd f+2\right)\pmbmt{\Lambda}_0^{-1}\pmbmt{A}_1^{-1}\right], \nonumber
\end{eqnarray}
\begin{eqnarray}
\pmbmt{{\cal{E}}}_{14}=\frac{\hat{f}}{\hat{\rho}x}V_Gc_1\bar{\omega}^2_A\left(\pmbmt{\Lambda}_0^{-1}\pmbmt{A}_0-m^2\pmbmt{\Lambda}_0^{-1}\pmbmt{A}_1^{-1}\right),
\quad \pmbmt{{\cal{E}}}_{15}=\frac{\hat{f}}{2\hat{\rho}x}V_Gc_1\bar{\omega}^2_Am\left[\frac{1}{2}\left(\rd f+2\right)\pmbmt{I}+2\pmbmt{\Lambda}_0^{-1}+\rd f\pmbmt{\Lambda}_0^{-1}\pmbmt{A}_1^{-1}\pmbmt{B}_1\pmbmt{\Lambda}_0+\pmbmt{\Lambda}_0^{-1}\pmbmt{A}_1^{-1}\tilde{\pmbmt{C}}_1\right], \nonumber
\end{eqnarray}
\begin{eqnarray}
\pmbmt{{\cal{E}}}_{16}=\frac{\hat{f}}{2\hat{\rho}x}V_Gc_1\bar{\omega}_A^2\pmbmt{\Lambda}_0^{-1}\pmbmt{A}_0\pmbmt{L}_0, \nonumber
\end{eqnarray}
\begin{eqnarray}
\pmbmt{{\cal{E}}}_{21}=\left(\rd f+2\right)\pmbmt{A}_1^{-1}\pmbmt{B}_1,\quad \pmbmt{{\cal{E}}}_{22}=-m\pmbmt{A}_1^{-1}, \quad \pmbmt{{\cal{E}}}_{23}=\frac{x}{2\hat{f}}\pmbmt{A}_1^{-1}, \nonumber
\end{eqnarray}
\begin{eqnarray}
\pmbmt{{\cal{E}}}_{31}=\left[\rd^2f+\left(2+V_G\right)\rd f-2\left(1-V_G\right)\right]\pmbmt{L}_1^{-1}\left(\pmbmt{A}_1^{-1}\pmbmt{B}_1+\frac{1}{2}m^2\pmbmt{A}_1^{-1}\pmbmt{A}_0^{-1}\right), \nonumber
\end{eqnarray}
\begin{eqnarray}
\pmbmt{{\cal{E}}}_{32}=-m\left(\rd f+2\right)\pmbmt{L}_1^{-1}\left(\pmbmt{A}_1^{-1}+\frac{1}{2}\pmbmt{A}_1^{-1}\pmbmt{A}_0^{-1}\tilde{\pmbmt{C}}_0\right), \quad \pmbmt{{\cal{E}}}_{34}=\frac{x}{2\hat{f}}\pmbmt{L}_1^{-1}\pmbmt{A}_1^{-1}\pmbmt{\Lambda}_1, \quad \pmbmt{{\cal{E}}}_{35}=-\frac{x}{2\hat{f}}m\pmbmt{L}_1^{-1}\pmbmt{A}_1^{-1}\pmbmt{A}_0^{-1}\pmbmt{\Lambda}_0, \nonumber
\end{eqnarray}
\begin{eqnarray}
\pmbmt{{\cal{E}}}_{37}=\left(\rd
  f+2\right)\pmbmt{L}_1^{-1}\left(m^2\pmbmt{A}_1^{-1}\pmbmt{A}_0^{-1}+\frac{1}{2}\pmbmt{A}_1^{-1}\tilde{\pmbmt{C}}_1\right),
\quad \pmbmt{{\cal{E}}}_{38}=-\left(\rd
  f+2\right)\pmbmt{L}_1^{-1}\left(\pmbmt{A}_1^{-1}\pmbmt{B}_1+\frac{1}{2}m^2\pmbmt{A}_1^{-1}\pmbmt{A}_0^{-1}\right),
\end{eqnarray}
\begin{eqnarray}
\pmbmt{{\cal{F}}}_{11}=\left(V_G-3\right)\pmbmt{I}, \quad \pmbmt{{\cal{F}}}_{21}=-m\left[\rd^2f+\left(2+V_G\right)\rd f-2\left(1-V_G\right)\right]\pmbmt{L}_0^{-1}\left(\pmbmt{A}_0^{-1}\pmbmt{A}_1^{-1}\pmbmt{B}_1+\frac{1}{2}\pmbmt{A}_0^{-1}\right), \nonumber
\end{eqnarray}
\begin{eqnarray}
\pmbmt{{\cal{F}}}_{22}=\left(\rd f+2\right)\pmbmt{L}_0^{-1}\left(m^2\pmbmt{A}_0^{-1}\pmbmt{A}_1^{-1}+\frac{1}{2}\pmbmt{A}_0^{-1}\tilde{\pmbmt{C}}_0\right),\quad \pmbmt{{\cal{F}}}_{24}=-\frac{x}{2\hat{f}}m\pmbmt{L}_0^{-1}\pmbmt{A}_0^{-1}\pmbmt{A}_1^{-1}\pmbmt{\Lambda}_1,\quad \pmbmt{{\cal{F}}}_{25}=\frac{x}{2\hat{f}}\pmbmt{L}_0^{-1}\pmbmt{A}_0^{-1}\pmbmt{\Lambda}_0, \nonumber
\end{eqnarray}
\begin{eqnarray}
\pmbmt{{\cal{F}}}_{52}=-2\frac{\hat{\rho}x}{\hat{f}}\frac{\bar{\sigma}^2}{\bar{\omega}_A^2}\pmbmt{A}_1^{-1}\pmbmt{\Lambda}_1,\quad
\pmbmt{{\cal{F}}}_{53}=\frac{1}{2}\left(\rd^2f+4\rd
  f+2\right)m\pmbmt{A}_1^{-1},\quad
\pmbmt{{\cal{F}}}_{54}=-2m\pmbmt{A}_1^{-1}, \quad
\pmbmt{{\cal{F}}}_{55}=\pmbmt{A}_1^{-1}\left(\rd
  f\pmbmt{B}_1\pmbmt{\Lambda}_0+\tilde{\pmbmt{C}}_1\right), \nonumber
\end{eqnarray}
\begin{eqnarray}
\pmbmt{{\cal{F}}}_{56}=-m\pmbmt{A}_1^{-1}, \quad \pmbmt{{\cal{F}}}_{61}=2\frac{\hat{\rho}x}{\hat{f}}\frac{1}{c_1\bar{\omega}_A^2}\left(c_1\bar{\sigma}^2+rA\right)\pmbmt{L}_0^{-1}\pmbmt{A}_0^{-1}\pmbmt{\Lambda}_0,\quad \pmbmt{{\cal{F}}}_{62}=4\frac{\hat{\rho}x}{\hat{f}}\frac{\bar{\sigma}^2}{\bar{\omega}_A^2}m\pmbmt{L}_0^{-1}\pmbmt{A}_0^{-1}\pmbmt{A}_1^{-1}\pmbmt{\Lambda}_1, \nonumber
\end{eqnarray}
\begin{eqnarray}
\pmbmt{{\cal{F}}}_{63}=\pmbmt{L}_0^{-1}\bigg\{-\frac{1}{2}\left(\rd f+2\right)\pmbmt{A}_0^{-1}\tilde{\pmbmt{C}}_0-\left(\rd\rho+4-rA\right)\pmbmt{I}+\left[-\rd^3f+\rd^2f\left(\rd\rho-4-rA\right)+4\rd f\left(\rd\rho+1-rA\right)\right]\pmbmt{A}_0^{-1}\pmbmt{B}_0 \nonumber \\
-\frac{1}{2}\left[\rd^3f-\rd^2f\left(\rd\rho-4\right)-2\rd f\left(1+2\rd\rho\right)-2\left(2+\rd\rho\right)\right]m^2\pmbmt{A}_0^{-1}\pmbmt{A}_1^{-1}\bigg\}, \nonumber
\end{eqnarray}
\begin{eqnarray}
\pmbmt{{\cal{F}}}_{64}=\pmbmt{L}_0^{-1}\bigg[\left(\rd f+1\right)\pmbmt{A}_0^{-1}\tilde{\pmbmt{C}}_0+\pmbmt{\Lambda}_1+2\pmbmt{A}_0^{-1}\pmbmt{B}_0\pmbmt{\Lambda}_1+2\left(\rd\rho-rA\right)\pmbmt{I}-\frac{1}{2}\left(\rd^2f+4\rd  f+2\right)m^2\pmbmt{A}_0^{-1}\pmbmt{A}_1^{-1}\tilde{\pmbmt{\Lambda}}_1 \nonumber \\
-\left(\rd^2f+2\rd f+2\rd\rho\right)m^2\pmbmt{A}_0^{-1}\pmbmt{A}_1^{-1}\bigg], \nonumber
\end{eqnarray}
\begin{eqnarray}
\pmbmt{{\cal{F}}}_{65}=\pmbmt{L}_0^{-1}\bigg\{-m\left\{2\left(\rd f-\rd\rho+1+rA\right)\pmbmt{A}_0^{-1}+\frac{1}{2}\left[-\rd f\left(\rd\rho+2\right)+\left(\rd f+2\right)rA-2\left(\rd\rho+1\right)\right]\pmbmt{A}_0^{-1}\pmbmt{\Lambda}_0\right\} \nonumber \\
-m\left[\left(\rd^2f-\rd\rho\rd f+2\rd f\right)\pmbmt{A}_0^{-1}\pmbmt{A}_1^{-1}\pmbmt{B}_1\pmbmt{\Lambda}_0+\left(\rd f-\rd\rho+1\right)\pmbmt{A}_0^{-1}\pmbmt{A}_1^{-1}\tilde{\pmbmt{C}}_1\right]\bigg\}, \nonumber
\end{eqnarray}
\begin{eqnarray}
\pmbmt{{\cal{F}}}_{66}=\pmbmt{L}_0^{-1}\left[\frac{1}{2}\left(\rd f+2\right)\pmbmt{A}_0^{-1}\tilde{\pmbmt{C}}_0+\left(\rd\rho-1-rA\right)\pmbmt{I}+\left(\rd f-\rd\rho+3\right)m^2\pmbmt{A}_0^{-1}\pmbmt{A}_1^{-1}\right],
\end{eqnarray}
\begin{eqnarray}
\pmbmt{{\cal{G}}}_{21}=m\left(\rd
  f+2\right)\pmbmt{L}_0^{-1}\left(\pmbmt{A}_0^{-1}\pmbmt{A}_1^{-1}\pmbmt{B}_1+\frac{1}{2}\pmbmt{A}_0^{-1}\right),
\quad \pmbmt{{\cal{G}}}_{22}=-m\left(\rd f+2\right)\pmbmt{L}_0^{-1}\left(\pmbmt{A}_0^{-1}+\frac{1}{2}\pmbmt{A}_0^{-1}\pmbmt{A}_1^{-1}\tilde{\pmbmt{C}}_1\right), \nonumber
\end{eqnarray}
\begin{eqnarray}
\pmbmt{{\cal{G}}}_{62}=-2\frac{\hat{\rho}x}{\hat{f}}\frac{\bar{\sigma}^2}{\bar{\omega}^2_A}\left(2+rA\right)\pmbmt{L}_0^{-1}\pmbmt{A}_0^{-1}\pmbmt{\Lambda}_0,\quad \pmbmt{{\cal{G}}}_{63}=-2\frac{\hat{\rho}x}{\hat{f}}\frac{\bar{\sigma}^2}{\bar{\omega}^2_A}\pmbmt{L}_0^{-1}\pmbmt{A}_0^{-1}\pmbmt{\Lambda}_0,\quad \pmbmt{{\cal{G}}}_{64}=2\frac{\hat{\rho}x}{\hat{f}}\frac{\bar{\sigma}^2}{\bar{\omega}^2_A}m\pmbmt{L}_0^{-1}\pmbmt{A}_0^{-1}\pmbmt{A}_1^{-1}\pmbmt{\Lambda}_1, \nonumber
\end{eqnarray}
\begin{eqnarray}
\pmbmt{{\cal{G}}}_{65}=-m\pmbmt{L}_0^{-1}\left[\frac{1}{2}\left(\rd f+2\right)\pmbmt{A}_0^{-1}\pmbmt{A}_1^{-1}\tilde{\pmbmt{C}}_1+\left(\rd f+2+rA\right)\pmbmt{A}_0^{-1}\right],
\end{eqnarray}
where
\begin{eqnarray}
V_G=\frac{V}{\Gamma_1},\quad x=\frac{r}{R},\quad
\hat{f}=\frac{f(r)}{f(0)},\quad
\hat{\rho}=\frac{\rho(r)}{\rho(0)},\quad
\omega_A^2=\frac{f^2(0)}{\pi\rho(0)R^2},\quad \beta=\frac{f^2(0)}{4\pi
  p}.
\end{eqnarray}

Using (A17)-(A25), we can summarize our master equations as follows:
\begin{eqnarray}
r\dfn{}{r}{}\left(\begin{array}{c}
\pmb{y}_1 \\
\pmb{y}_2 \\
\pmb{y}_3 \\
\pmb{y}_4 \\
\pmb{y}_5 \\
\pmb{y}_6
\end{array}\right)=\pmbmt{{\cal{A}}}\left(\begin{array}{c}
\pmb{y}_1 \\
\pmb{y}_2 \\
\pmb{y}_3 \\
\pmb{y}_4 \\
\pmb{y}_5 \\
\pmb{y}_6
\end{array}\right), 
\end{eqnarray}
where $\pmbmt{{\cal{A}}}$ is a matrix derived through complicated calculations. 
We assume that no electric current flows outside the star. 
Therefore, the magnetic field perturbation regular at infinity needs to satisfy the relations, given by 
\begin{eqnarray}
\pmb{h}^S+{\bf L^+} \pmb{h}^H=0 \,, \quad \pmb{h}^T=0 \,, \quad {\rm for} \ r > R \,, 
\label{bc-mag}
\end{eqnarray}
where $({\bf L^+} )_{ij}=(l_j^\prime+1)\delta_{ij}$. 
The surface boundary conditions assumed in this study are given by 
\begin{eqnarray}
[[\Delta {\bf B}]]=0 \,, \quad \Delta p=0 \,, 
\end{eqnarray}
where for the physical quantity $Q$, $\Delta Q$ denotes the Lagrangian change in $Q$ and 
$\displaystyle [[Q]]\equiv\lim_{\epsilon\rightarrow 0}\{ Q(R-\epsilon)-Q(R+\epsilon)\}$ 
for $\epsilon >0$. Note that the mechanical surface boundary condition is usually given by 
$\displaystyle \left[\left[\Delta \left(p+{1\over 8\pi}|{\bf B}|^2\right)\right]\right]=0$. However, the second term in the left-hand side of this 
equation automatically vanishes due to 
the conditions of $[[\Delta {\bf B}]]=0$ and $[[{\bf B}]]=0$, which are assumed in this study.   
For the magnetized star models employed in this study, the electric current 
vanishes at the surface of the star. We then have $[[\Delta {\bf B}]]=[[{\bf B}']]=0$, from which Eq. (\ref{bc-mag}) has to be satisfied  
at the surface of the star. The condition $\Delta p=0$ is explicitly written by 
\begin{eqnarray}
\delta\pmb{U}-\pmb{y}_1=0\,. 
\end{eqnarray}
The boundary conditions at the stellar center are
the regularity conditions for the eigenfunctions
$\pmb{y}_1$-- $\pmb{y}_6$. We adopt a normalization condition
$T_{l_1^\prime}(R)=1$ at the stellar surface.



\label{lastpage}

\end{document}